\documentclass[prl,twocolumn,showpacs,amsmath,amssymb,superscriptaddress]{revtex4-1}
\usepackage{dcolumn}
\usepackage{bm,graphicx}
\usepackage{url}
\usepackage{color}

\usepackage[T1]{fontenc} 
\usepackage[english]{babel} 
\usepackage{lmodern} 
\usepackage{amsmath,amsfonts,amssymb} 
\usepackage{xcolor} 
\usepackage{url} 
\usepackage{subfigure} 
\usepackage{blindtext} 
\usepackage{lipsum}

\newcommand{\ket}[1]{\ensuremath{\left|#1\right\rangle}}

\addto\captionsngerman 
{ }
\addto\captionsenglish 
{} 

\renewcommand{\vec}[1]{\boldsymbol{#1}} 
\newcommand{\betrag}[1]{\left\vert #1 \right\vert} 
\newcommand{\bra}[1]{\left\langle #1 \right\vert} 

\begin{document}

\title{Magneto-optics of  massive Dirac fermions in bulk Bi$_2$Se$_3$}

\author{M. Orlita}\email{milan.orlita@lncmi.cnrs.fr}
\affiliation{Laboratoire National des Champs Magn\'etiques Intenses, CNRS-UJF-UPS-INSA, 25, avenue des
Martyrs, 38042 Grenoble, France}
\author{B. A. Piot}
\affiliation{Laboratoire National des Champs Magn\'etiques Intenses,
CNRS-UJF-UPS-INSA, 25, avenue des Martyrs, 38042 Grenoble, France}
\author{G. Martinez}
\affiliation{Laboratoire National des Champs Magn\'etiques Intenses,
CNRS-UJF-UPS-INSA, 25, avenue des Martyrs, 38042 Grenoble, France}
\author{N. K. Sampath Kumar}
\affiliation{Laboratoire National des Champs Magn\'etiques Intenses,
CNRS-UJF-UPS-INSA, 25, avenue des Martyrs, 38042 Grenoble, France}
\author{C. Faugeras}
\affiliation{Laboratoire National des Champs Magn\'etiques Intenses,
CNRS-UJF-UPS-INSA, 25, avenue des Martyrs, 38042 Grenoble, France}
\author{M.~Potemski}
\affiliation{Laboratoire National des Champs Magn\'etiques Intenses,
CNRS-UJF-UPS-INSA, 25, avenue des Martyrs, 38042 Grenoble, France}
\author{C.~Michel}
\affiliation{Institute for Theoretical Physics, TP IV, University of W\"{u}rzburg, Am Hubland, D-97074 W\"{u}rzburg, Germany}
\author{E.~M.~Hankiewicz}
\affiliation{Institute for Theoretical Physics, TP IV, University of W\"{u}rzburg, Am Hubland, D-97074 W\"{u}rzburg, Germany}
\author{T.~Brauner}
\affiliation{Institute for Theoretical Physics, Vienna University of Technology, 
A-1040 Vienna, Austria}
\author{\v{C}. Dra\v{s}ar}
\affiliation{
Faculty of Chemical Technology, University of Pardubice,
CZ-53210 Pardubice, Czech Republic}
\author{S.~Schreyeck}
\affiliation{Physikalisches Institut (EP III), Universit\"{a}t W\"{u}rzburg, D-97074 W\"{u}rzburg, Germany}
\author{S.~Grauer}
\affiliation{Physikalisches Institut (EP III), Universit\"{a}t W\"{u}rzburg, D-97074 W\"{u}rzburg, Germany}
\author{K.~Brunner}
\affiliation{Physikalisches Institut (EP III), Universit\"{a}t W\"{u}rzburg, D-97074 W\"{u}rzburg, Germany}
\author{C.~Gould}
\affiliation{Physikalisches Institut (EP III), Universit\"{a}t W\"{u}rzburg, D-97074 W\"{u}rzburg, Germany}
\author{C.~Br\"{u}ne}
\affiliation{Physikalisches Institut (EP III), Universit\"{a}t W\"{u}rzburg, D-97074 W\"{u}rzburg, Germany}
\author{L.~W.~Molenkamp}
\affiliation{Physikalisches Institut (EP III), Universit\"{a}t W\"{u}rzburg, D-97074 W\"{u}rzburg, Germany}
\date{\today}

\begin{abstract}
We report on magneto-optical studies of Bi$_2$Se$_3$, a
representative member of the 3D topological insulator family. Its
electronic states in bulk are shown to be well described by a
simple Dirac-type Hamiltonian for massive particles with only two parameters:
the fundamental bandgap and the band velocity.
In a magnetic field, this model implies a unique property -- spin splitting equal to twice the cyclotron
energy: $E_s=2E_c$. This explains the extensive
magneto-transport studies concluding a fortuitous degeneracy
of the spin and orbital split Landau levels in this material.
The $E_s=2E_c$ match differentiates the massive Dirac electrons
in bulk Bi$_2$Se$_3$ from those in quantum electrodynamics,
for which $E_s=E_c$ always holds.

\end{abstract}


\maketitle

Inspiring analogies to relativistic systems have largely helped to
elucidate the electronic properties of two-dimensional
graphene~\cite{NovoselovNature05,ZhangNature05}, surface states of
topological insulators
(TIs)~\cite{BernevigPRL06,*KonigScience07,HsiehNature08,ZhangNaturePhys09,HasanRMP10,*QiRMP11},
novel three-dimensional (3D)
semimetals~\cite{LiuScience14,OrlitaNaturePhys14,LiuNatureMater14}
as well as certain narrow gap
semiconductors~\cite{ZawadzkiAinP74}. Here, we report on
magneto-optical studies of bulk Bi$_2$Se$_3$, which
imply the approximate applicability of the Dirac Hamiltonian for massive relativistic particles
to approach the band structure of this popular representative of the TI family.

The dispersion relations of genuine massive Dirac fermions
in quantum electrodynamics are defined by two parameters:
the energy gap $2\Delta$ between particles and antiparticles and velocity
parameter $v_D$. At low energies, \textit{i.e.}, in the non-relativistic
limit, these dispersions become parabolic and characterized by
the same effective mass $m_D=\Delta/v_D^2$ (rest Dirac mass).
Such dispersions resemble the cartoon sketch of a direct
gap semiconductor, which may be conventionally described
using Schr\"{o}dinger equation, completed by extra Pauli terms
in order to include the spin degree of freedom. In contrast,
no additional terms are needed when Dirac equation is employed,
since it inherently accounts for spin-related effects.
For instance, when the magnetic field $B$ is applied, Dirac equation describes both
cyclotron ($E_c$) as well as spin ($E_s$) splitting of
the electronic states and implies that these two splitting energies
are the same and linear with $B$ in the
non-relativistic approximation: $E_s =
E_c=\hbar\omega_c = \hbar eB/m_D = e\hbar B v_D^2/\Delta$.
For free electrons, this condition is equivalent to the effective $g$
factor of 2 in $E_s=g\mu_B B$ (Bohr magneton $\mu_B=e\hbar/2m_0$)~\cite{Landau77}.

In this Letter, we demonstrate experimentally that the conduction
and valence bands of Bi$_2$Se$_3$ are both, with a good precision,
parabolic (perpendicular to the $c$-axis) and
characterized by approximately the same effective mass. This
crucial observation implies a great simplification of the
multi-parameter Dirac Hamiltonian~\cite{ZhangNaturePhys09,LiuPRB10}
commonly used to describe the bands of this material.

The resulting simplified Dirac Hamiltonian differs from that of the genuine
quantum electrodynamics system only by relevant (additional) diagonal
dispersive terms, and importantly, it remains to be defined by two parameters only:
by the bandgap energy $2\Delta$ and velocity parameter $v_D$.
These are directly read from our optical experiments, or alternatively,
the $v_D$ parameter may be taken from the
measurements of the Bi$_2$Se$_3$ Dirac-cone surface states~\cite{XiaNaturePhys09}.
Remarkably and in contrast to genuine Dirac fermions,
the electrons in Bi$_2$Se$_3$ approximately follow the rule that their spin splitting
is twice the cyclotron energy $E_s=\hbar e B
v_D^2/\Delta=2\hbar\omega_c=2E_c$. The effective mass, common
for carriers in the conduction and valence bands, is thus roughly
$m_e=m_h= 2\Delta/v_D^2=2m_D$ and the spin splitting
expressed in terms of the effective $g$ factor, $g_e=g_h=2m_0/m_D$. Our
simplified view of the bands in Bi$_2$Se$_3$ is not perfect
(departures are extensively discussed), though it accounts well for
the present experimental results as well as for a number of
magneto-transport data reported in the past and agrees with the recent
estimate of the electron $g$ factor.

\begin{figure}[t]
      \includegraphics[trim = 0mm 0mm 45mm 125mm, clip=true, width=8.5cm]{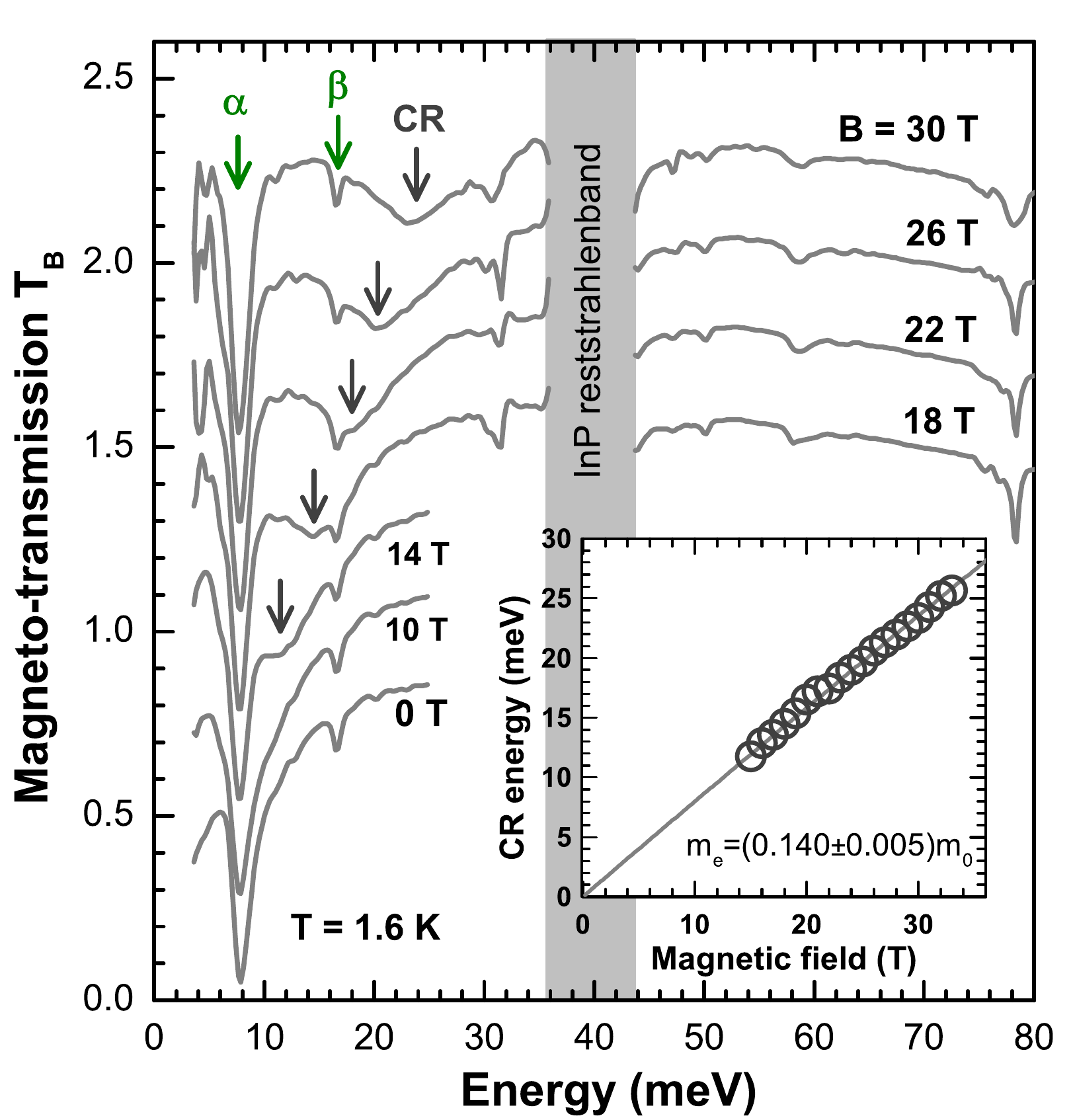}
      \caption{\label{CR} (color online) Far infrared magneto-transmission spectra of the Bi$_2$Se$_3$ specimen. The infrared active phonon modes $\alpha$ and $\beta$
      are at higher magnetic fields accompanied by CR absorption, which follows linear in $B$ dependence, see the inset, and
      implies the effective mass of electrons $m_e=(0.140\pm0.005)m_0$. The plotted transmission $T_B$ was
      normalized by that of the bare InP substrate.}
\end{figure}

\begin{figure*}
    \begin{minipage}{0.65\linewidth}
    \includegraphics[trim = 10mm 10mm 10mm 160mm, clip=true, width=12cm]{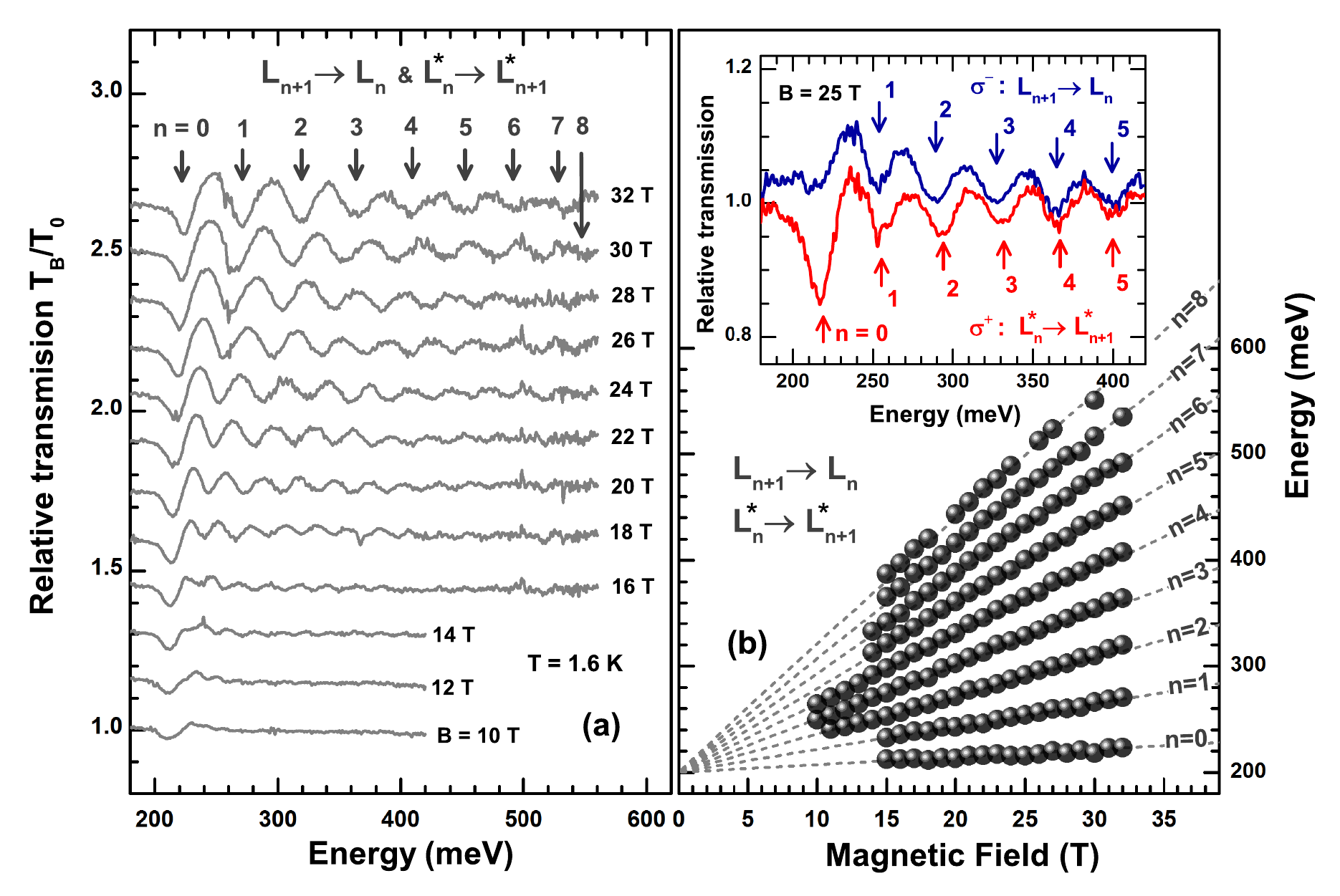}
    \end{minipage}\hfill
    \begin{minipage}{0.32\linewidth}
      \caption{\label{Interband_highB} (color online) Part (a): Relative transmission spectra of Bi$_2$Se$_3$ in the middle infrared spectral range plotted for selected values of $B$.
      At $B=32$~T, individual excitations are denoted by vertical arrows and identified by the corresponding index $n$.
      Part (b): Positions of experimentally observed interband excitations as a function of $B$. The dashed lines represent theoretical
      fit to data described in the text. The inset shows the magneto-transmission spectrum taken at
      $B=25$~T measured with a defined circular polarization of light. Notably, it is the normalization by $T_0$,
      which induces the modulation of $T_B/T_0$ curves around the zero-field interband absorption edge of $\hbar\omega=2\Delta+E_F(1+m_e/m_h)\approx225$~meV.}
    \end{minipage}
\end{figure*}

The presented experiments have been performed on a 290-nm-thick layer of Bi$_2$Se$_3$ grown by molecular beam epitaxy
on a semi-insulating InP(111)B substrate~\cite{TarakinaAMI14}; for data obtained on another specimen prepared
under analogous conditions see Supplementary materials~\cite{SI}. The 3D structure of the InP surface
(with 2~nm root mean square roughness) transfers the stacking order of the substrate to the epilayer, resulting in a complete suppression
of twinning, which is otherwise a ubiquitous defect in Bi$_2$Se$_3$ thin films. The after-growth annealing in the Se atmosphere, reducing the final density of Se vacancies,
helped to keep the electron density below $10^{18}$~cm$^{-3}$ (with the mobility $\mu$ in the $10^3$~cm$^2$.V$^{-1}$.s$^{-1}$range), as confirmed in magneto-transport experiments.
Importantly, the thin epitaxial layer of Bi$_2$Se$_3$ enabled transmission experiments at photon
energies above the fundamental interband absorption edge of this
material.

To measure the magneto-transmission spectra, a macroscopic area of
the sample ($\sim$4~mm$^2$) was exposed to the radiation of a
globar, which was analysed by a Fourier transform spectrometer
and, using light-pipe optics, delivered to the sample placed in a
superconducting or resistive magnet. The transmitted light was
detected by a composite bolometer placed directly below the
sample, kept at a temperature of 1.6~K. All measurements were done
in the Faraday configuration with light propagating along the $c$
axis of Bi$_2$Se$_3$ ($z$ axis). In experiments performed
with circularly polarized light, a glass linear polarizer and a
zero-order MgF$_2$ quarter wave plates (centered at $\lambda = 4$
or 5~$\mu$m) were used.

The optical response of the thin Bi$_2$Se$_3$ layer has been
probed in both far and middle infrared spectral regions. At low
energies, the response, see Fig.~\ref{CR}, is dominated by
infrared active phonon modes $\alpha$ and $\beta$, which exhibit a
weak coupling to the magnetic field~\cite{LaForgePRB10}. At higher
magnetic fields, cyclotron resonance (CR) absorption is well
formed and it disperses linearly with $B$. The slope of this
dependence provides us with an estimate of the electron effective mass:
$m_e=(0.140\pm0.005)m_0$, which well falls into a relatively broad
range of values, $m_e=(0.12-0.16)m_0$, deduced from
other experiments~\cite{KohlerPSSB75,*EtoPRB10,AnalytisPRB10,ButchPRB10,SushkovPRB10,CaoPRL12}.
The interband absorption of Bi$_2$Se$_3$ exhibits a fairly rich
response in magnetic fields, see Fig.~\ref{Interband_highB}(a).
Firstly, at low $B$, a distortion of the absorption
edge at the energy slightly above 200~meV appears. At higher
fields ($B>10$~T), the quantum regime is approached ($\mu.B>1$)
and a series of interband inter-Landau level (inter-LL)
resonances emerges. These resonances are
almost equidistant in energy and follow nearly linear in $B$
dependence, see Fig.~\ref{Interband_highB}(b).

The observed linearity of the optical response in $B$, in
reference to intraband (CR absorption) as well as interband
inter-LL excitations, points towards parabolic profiles of both
conduction and valence bands. Let us reconcile this crucial
experimental fact with the standard theoretical model of
electronic bands in TIs from Bi$_2$Se$_3$ family~\cite{ZhangNaturePhys09,LiuPRB10}.
Using a basis of spin-degenerate Se- and Bi-like $p$-orbitals,
the authors of Refs.~\cite{ZhangNaturePhys09,LiuPRB10} propose a 3D Dirac Hamiltonian ($4\times4$),
expanded to include the electron-hole
asymmetry, uniaxial anisotropy (along the $c$ axis), and
importantly, the band inversion, giving thus rise to the
TI phase (via dispersive diagonal terms).

Since these are the $k_z=0$ states, which provide the dominant
contribution to the magneto-optical response studied in our
experiments, the situation further simplifies. The 3D Dirac
Hamiltonian decouples into two complex-conjugate 2D Dirac-type
Hamiltonians $h$ and $h^*$ written in the basis of
$\ket{\mathrm{Se}\downarrow}, \ket{\mathrm{Bi}\uparrow}$ and
$\ket{\mathrm{Se}\uparrow}, \ket{\mathrm{Bi}\downarrow}$,
respectively:
\begin{equation}\label{Hamiltonian}
h=\left(\begin{array}{cc}\Delta +(C+M)k^2 &  \hbar v_D k_+\\
\hbar v_D k_- & -\Delta+(C-M) k^2\\
\end{array}\right),
\end{equation}
where $k_\pm=k_x\pm i k_y$. This Hamiltonian implies, in general, non-parabolic conduction and valence band profiles:
\begin{equation}\label{dispersion}
\mathcal{E}_{c,v}(k)=Ck^2\pm\sqrt{(\Delta+M k^2)^2+\hbar^2 v_D^2k^2},
\end{equation}
each exhibiting up to three local extremal points, depending on
the strength of the interband coupling $v_D$ (effective speed of
light), the electron-hole asymmetry parameter $C$, and the
diagonal dispersive term $M$ (negative for systems with the band
inversion). The basis of the $h$ Hamiltonian allows us to
associate a given spin projection to each band:
$\mathcal{E}_{c\downarrow}$ and $\mathcal{E}_{v\uparrow}$. To
satisfy the time-reversal and inversion symmetries of
Bi$_2$Se$_3$, the Hamiltonian $h^*$ provides an analogous
solution, with the spin projections rotated,
$\mathcal{E}_{c\uparrow}$ and $\mathcal{E}_{v\downarrow}$, so we
finally obtain twice spin-degenerate conduction and valence bands.

To make the above dispersions \eqref{dispersion} parabolic in a
broad range of energies, \emph{i.e.}, to make the model consistent
with our magneto-optical data, the specific condition $\hbar^2
v_D^2=-4M\Delta$ has to be satisfied. Notably, this is only
possible for systems in the TI phase when $M<0$ (by definition $\Delta>0$). The bands then take a
simple form, $\mathcal{E}_c=\Delta+(C-M)k^2$ and
$\mathcal{E}_v=-\Delta+(C+M)k^2$, and are characterized by
well-defined effective masses:
$m_e=\hbar^2/[2(C-M)]=2\hbar^2/(\hbar^2/m_D+4C)$ and
$m_h=-\hbar^2/[2(C+M)]=2\hbar^2/(\hbar^2/m_D-4C)$, for electrons and
holes, respectively. Interestingly, the corresponding reduced mass
equals to the Dirac mass: $1/m_e + 1/m_h = 1/m_D = v_D^2/\Delta $.
Clearly, in the case of a relatively weak electron-hole asymmetry
($C\ll |M|$), the expressions further reduce to $m_e\approx m_h
\approx 2m_D$.

\begin{figure*}
\includegraphics[trim = 12mm 12mm 12mm 200mm, clip=true, width=17cm]{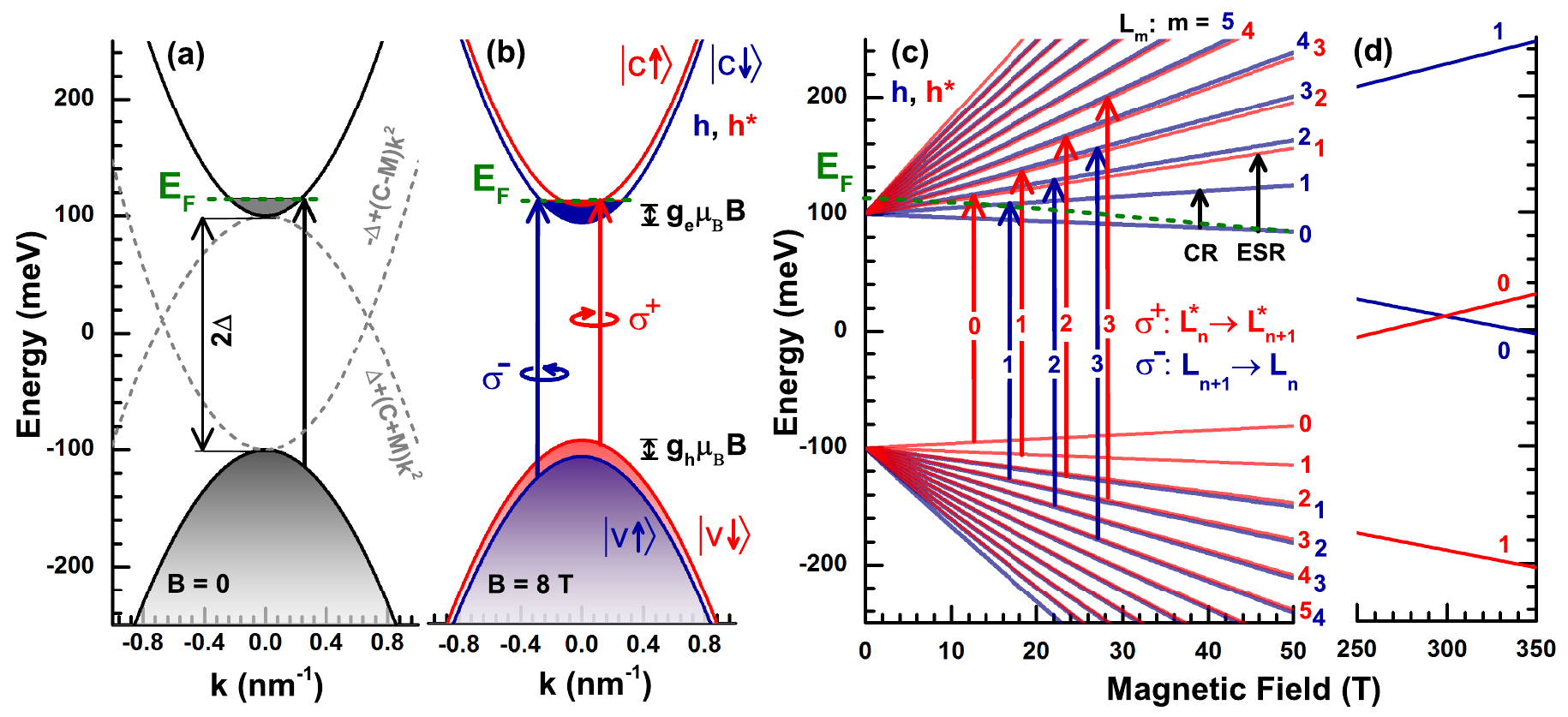}
      \caption{\label{Illustrative} (color online) Part (a): Approximate profile of electronic bands in Bi$_2$Se$_3$ at the center of the Brillouin zone (for $k_z=0$). The
      dashed lines show the (dispersive) diagonal term of the Hamiltonian \eqref{Hamiltonian}. Part (b): Spin splitting of the conduction and valence bands ($g_e\approx g_h$) at $B=8$~T
      (LLs not yet resolved $\mu.B<1$). Part (c): Fan chart of nearly linear in $B$ LLs in Bi$_2$Se$_3$. The approximate match of LLs $E_{n+1,e}\approx E^*_{n,e}$ \& $E_{n,h}\approx E^*_{n+1,h}$ is a direct consequence of the Dirac-type model for electronic states in Bi$_2$Se$_3$ discussed in the text. Vertical arrows show selected electric-dipole-active interband inter-LL resonances, CR absorption as well as (expected) magnetic-dipole-active electron spin resonance (ESR) absorption in the quantum limit. Part (d): High-magnetic-field extrapolation of zero-mode LLs, with the crossing point at $B_c\approx 300$~T.}
\end{figure*}

When the magnetic field is applied, the bands in
Bi$_2$Se$_3$ transform into Landau levels (LLs). The Dirac-type
Hamiltonians $h$ and $h^*$ give rise to particular electron and
hole zero-mode LLs: $E_{0,e} = \Delta+(C+M)eB/\hbar$ and
$E^*_{0,h} = -\Delta+(C-M)eB/\hbar$. These zero-mode levels are typical of
TIs (see, \textit{e.g.}, Refs.~\cite{BernevigPRL06,KonigScience07})
-- they disperse strictly linearly with $B$, they are spin
polarized, insensitive to the strength of the interband coupling
$v_D$ and they cross each other at the field of $B_c=\hbar
\Delta/|eM|$. The LLs with higher indices ($n>0$) follow, assuming
parabolic bands with a relatively weak electron-hole
asymmetry $C\ll |M|$ (a posteriori justified by our experimental data),
nearly linear in $B$ dependence. For $h$ and $h^*$ Hamiltonians we get the LL spectrum: $E_{n,e} =
E_{0,e}+\hbar\omega_c^e n$, $E^*_{n,e}=E_{n-1,e}+\hbar\omega_c^D$,
$E^*_{n,h} = E^*_{0,h}-\hbar\omega_c^h n$ and
$E_{n,h}=E^*_{n-1,h}-\hbar\omega_c^D$, where $\omega_c^{e,h,D}=eB/m_{e,h,D}$.

Importantly, the shift $\hbar\omega_c^D$ between the two LL series
corresponds to the spin splitting $E_s=\hbar eBv_D^2/\Delta$, which may
be expressed also in terms of a $g$ factor: $g_e=g_h=2m_0/m_D$,
see Supplementary materials~\cite{SI}. In analogy to massive
particles in quantum electrodynamics, this spin splitting is given just by the energy
bandgap ($2\Delta$) and the effective velocity of light ($v_D$)
and it is the same for electrons and holes (particles and
antiparticles). On the other hand, the effective masses of
electrons and holes depend on the diagonal terms $M$ and $C$ and
this implies a certain ratio, $E_s/E_c>1$, between the cyclotron
energy and spin splitting (notably, $E_c/E_s=1$ always holds
for free electrons in vacuum). The spin-splitting of electronic
bands in TIs (with $M<0$) should thus manifest at lower magnetic
fields, prior to Landau level quantization cf.
Figs.~\ref{Illustrative}(a-c). This, in fact, accounts for the
initial distortion of the absorption edge observed at low magnetic
fields, see Fig.~\ref{Interband_highB}(a). Interestingly, for rather small electron-hole asymmetry
($C\ll|M| \Rightarrow m_e\rightarrow 2m_D$), we get
$2E_c=2\hbar\omega_c=\hbar eB/m_D=g_e\mu_B B=E_s$.

The Dirac Hamiltonian \eqref{Hamiltonian} in a magnetic field
gives rise to two series of dipole-active inter-LL excitations
$n\,$$\rightarrow$$\,n+1$ and $n\,$$\rightarrow$$\,n-1$, active in
$\sigma^+$ and $\sigma^-$ polarized light, respectively. This has
been corroborated both experimentally and theoretically, for
instance, in the context of
graphene~\cite{SadowskiPRL06,GusyninPRL07}.
However, a closer look at the corresponding matrix elements shows
that, in a gapped system ($\Delta \neq 0$),
$n\,$$\rightarrow$$\,n-1$ transitions dominate interband inter-LL
absorption within the spectrum of the $h$
Hamiltonian, and vice versa, the $n\,$$\rightarrow$$\,n+1$ series
in the $h^*$ Hamiltonian, see Fig.~\ref{Illustrative}(c) for
illustration and Supplementary materials~\cite{SI} for details. This behavior may be viewed as a spin-dependent
optical activity, implying necessity to reverse spin during the
interband absorption. The interband excitations,
$\mathcal{E}_{v\downarrow}\,$$\rightarrow$$\,\mathcal{E}_{c\uparrow}$
and
$\mathcal{E}_{v\uparrow}\,$$\rightarrow$$\,\mathcal{E}_{c\downarrow}$,
connecting states within $h^*$ or $h$ Hamiltonians, respectively,
are thus active in $\sigma^+$ and $\sigma^-$ polarized radiation
only. Notably, a similar situation is encountered in gapped
graphene and transition-metal dichalcogenides, where
valley-sensitive selection rules for circularly polarized light
appear~\cite{YaoPRB08,*RosePRB13}. Let us also note that intraband
(CR) absorption is always active in $\sigma^+$ and $\sigma^-$
polarized radiation for electrons and holes, respectively.

The LL spectrum and the selection
rules allows us to identify individual resonances in the
interband response in Fig.~\ref{Interband_highB}. When the quantum limit is reached
($B>20$~T), the lowest in energy observed absorption line is the
L$^*_0\rightarrow\,$L$^*_1$ transition, active in $\sigma^+$
polarized light. The parent L$_1\rightarrow\,$L$_0$ line, active
in $\sigma^-$ polarized radiation, does not appear, since the
bottom of $E_0$ level is always occupied in the $n$ doped system,
cf. the inset of Fig.~\ref{Interband_highB}(b) and
Fig.~\ref{Illustrative}(c). At higher energies, we get a series of
transitions, L$^*_n\rightarrow\,$L$^*_{n+1}$ and
L$_{n+1}\rightarrow\,$L$_{n}$ ($n>0$), which are for a given $n$
nearly degenerate in energy and active in $\sigma^+$ and
$\sigma^-$ polarized light, respectively. Their spacing,
$\hbar(\omega_c^e+\omega_c^h)=\hbar \omega_c^D$, allows us to read
the Dirac mass directly from the data: $m_D=(0.080 \pm 0.005)m_0$.
Notably, this Dirac mass implies $g_e=g_h=2m_0/m_D\approx25$, which is
in very good agreement with the value
$g_e^{\mathrm{ESR}}=27.5$~\cite{WolosAIP13}, derived recently
using spin resonance measurements. The Dirac mass $m_D$,
together with the bandgap $2\Delta=(200\pm5)$~meV, read from the
low-field extrapolation of interband inter-LL resonances in
Fig.~\ref{Interband_highB}(b), imply the velocity
$v_D=\sqrt{\Delta/m_D}=4.8\times10^5$~m/s, in perfect agreement with majority of
ARPES studies~\cite{XiaNaturePhys09,ZhuPRL11}.

Comparing the estimated Dirac mass with the electron mass (as
deduced from CR absorption), we conclude that $m_e \approx 2m_D$.
This indicates rather weak electron-hole asymmetry in Bi$_2$Se$_3$
($C\ll|M|=\hbar^2v_D^2/4\Delta$). Neglecting this asymmetry
completely, we remain with the Hamiltonian \eqref{Hamiltonian}
with two independent parameters only: $\Delta$ and $v_D$.
Interestingly, these two parameters, which can be easily read
from infrared transmission experiments, fully describe the band structure of Bi$_2$Se$_3$:
the energy bandgap of $2\Delta$, effective masses $m_e\approx
m_h\approx 2m_D = 2\Delta/v_D^2$, as well as $g$ factors,
$g_e=g_h=2m_0/m_D=2m_0v^2_D/\Delta$. Notably, this match between
twice cyclotron energy and spin splitting ($2\hbar e B/m_e=g_e\mu_B B$) has been in the
past found as a purely empirical fact in quantum oscillation
experiments on Bi$_2$Se$_3$, see, \textit{e.g.},
Refs.~\cite{KohlerPPS75,FauquePRB13}. Here we show that this
surprising match is not accidental and it
straightforwardly follows from the Dirac-type Hamiltonian
\eqref{Hamiltonian} applied to TI with a weak electron-hole
asymmetry and nearly parabolic bands.

It should be also mentioned that within our ``parabolic view'' of electronic bands,
Bi$_2$Se$_3$ clearly becomes a direct-gap semiconductor. This is in agreement with
recent experimental (ARPES) and theoretical studies, see, \textit{e.g.}, Refs.~\cite{ChenScience10,YazyevPRB12,AguileraPRB13,NechaevPRB13},
however, in contradiction with other ARPES data,
see, \textit{e.g.}, Refs.~\cite{XiaNaturePhys09,ZhuPRL11}, in which the observed
camel-back profile of the valence band indicated an indirect band gap. Intriguingly,
our experiments, together with other optical studies performed on bulk or thin-film specimens,
see, \textit{e.g.}, Refs.~\cite{GreenawayJPCS65,KohlerPPS74,PostPRB13},
provide a significantly lower band gap (200~meV) as compared to values deduced from ARPES experiments
($\sim$300~meV), see Supplementary materials~\cite{SI} for further discussions.

Now we will discuss the limits of our simplified two-parameter
model (with parabolic bands and full electron-hole symmetry),
by confronting it with more detailed analysis of our
experimental data. The real band structure of Bi$_2$Se$_3$
may deviate by (i) appearance
of the electron-hole asymmetry and (ii) the departure of bands
from exact parabolicity. The electron-hole asymmetry is clearly
demonstrated by $m_e<m_h$ ($m_e<2m_D$), which translates into
$C=(3\pm0.5)$~eV.\AA$^2$, but also by the difference in the
corresponding $g$ factors. The latter may be read from a small,
but noticeable, splitting between
L$^*_n\rightarrow\,$L$^*_{n+1}$ and L$_{n+1}\rightarrow\,$L$_n$
transitions in the spectra taken with a defined circular
polarization of light, see the inset of
Fig.~\ref{Interband_highB}(b). It implies $g_e-g_h\approx 3$,
which may be explained as the contribution of the free-electron
Zeeman term and influence of remote bands, described by,
\textit{e.g.}, the Roth's formula~\cite{RothPR59,LiuPRB10}.

The deviations from bands' parabolicity imply
the departure of inter-LL resonances from their linearity in
$B$. Indeed, the transitions at higher energies and/or for higher LL
indices, see Fig.~\ref{Interband_highB}(b), slow down to a weak sublinear dependence. To
describe this behavior, we have used the full (non-linearized)
expressions for LLs, see Supplementary materials~\cite{SI}, to fit
the positions of individual resonances. We varied parameters $v_D$,
$M$ and $\Delta$, while fixing $C\equiv0$, which has rather weak impact
on the interband response. The best agreement is
obtained for $v_D=(0.47\pm0.02)\times10^6$~m.s$^{-1}$,
$\Delta=(0.100\pm0.002)$~eV and $M=-(22.5\pm1.0)$~eV.\AA$^2$. We
may thus conclude that the condition $\hbar^2
v_D^2=-4M\Delta$ is fulfilled within a few percent, which
validates our view of parabolic bands in Bi$_2$Se$_3$.
Moreover, since $C/|M|\sim1/10$, the system indeed exhibits rather high electron-hole symmetry.

The deduced strength of dispersive diagonal terms, $M$ and
$\Delta$, allows us to estimate the critical field $B_c$, at which
the zero-mode LLs cross each other, see
Fig.~\ref{Illustrative}(d). At this magnetic field, Bi$_2$Se$_3$
changes into a semi-metallic (= gapless) material, for which the
extended 3D Dirac Hamiltonian implies, see Supplementary
materials~\cite{SI}, the linear in $k_z$ bands, $E(k_z)=\pm
\hbar\tilde{v}_D|k_z|$, with a high (LL) degeneracy
$\zeta=eB/\hbar$. The velocity $\tilde{v}_D$ is supposed to be
slightly lower as compared to $v_D$ due to the uniaxial anisotropy
of Bi$_2$Se$_3$~\cite{KohlerPSSB75,FauquePRB13}. The pretty
high value of $B_c\approx300$~T, at the limit of currently
available (semi-destructive) pulsed-field
techniques~\cite{PortugallCRP13}, makes the exploration of this
interesting critical point difficult.
However, this crossing field
$B_c$ is expected to be lower in other TIs from the Bi$_2$Se$_3$
family, with a lower band gap, \textit{e.g.}, in Bi$_{1-x}$Sb$_x$
for rather low Sb concentrations~\cite{HsiehNature08}.

In conclusion, we have shown that the band structure of
Bi$_2$Se$_3$ can be, in very good approximation, described by a
simple Dirac-type Hamiltonian with only two free parameters: the
effective velocity parameter $v_D$ and the band gap $2\Delta$.
This simplified model provides us with reasonable estimates for
both effective masses ($m_e\approx m_h\approx2\Delta/v^2_D$) and
corresponding $g$ factors ($g_e\approx g_h\approx2m_0
v_D^2/\Delta$), and implies, for charge carriers in Bi$_2$Se$_3$,
a surprising match between the cyclotron energy and
spin-splitting: $E_s\approx 2E_c$. Notably, this relation has been
deduced from quantum oscillations experiments performed on
Bi$_2$Se$_3$ in the past, but only as a purely empirical fact.
Here we show that this directly follows from the Dirac-type
Hamiltonian applied to a TI with nearly
parabolic bands and a high electron-hole symmetry.

\begin{acknowledgments}
The work has been supported by the ERC-AG projects MOMB and 3-TOP. Authors
acknowledge discussions with D. M. Basko, O. Ly and M. O. Goerbig.
T. B. acknowledges the support from the Austrian Science
Fund (FWF), Grant No. M 1603-N27. E. M. H. and and C. M. thank DFG grant HA 5893/4-1 within SPP 1666.
\end{acknowledgments}


%

\newpage

\vspace{2cm}

\begin{widetext}
\begin{center}
{\large
Supplementary Information for\\\vspace{2mm}
\vspace{0.2cm}
\textbf{Magneto-optics of  massive Dirac fermions in bulk Bi$_2$Se$_3$}\\\vspace{2mm}
\vspace{0.2cm}
by M. Orlita, B. A. Piot, G. Martinez, N. K. Sampath Kumar, C. Faugeras,
M. Potemski, \newline C. Michel, E. M. Hankiewicz, T. Brauner, \v{C}. Dra\v{s}ar,\newline S. Schreyeck,
S. Grauer, K. Brunner, C. Gould, C. Br\"{u}ne, and L. W. Molenkamp}
\end{center}

\vspace{0.5cm}
In this supplementary material, we present details of Landau level (LL) spectrum and optical selection rules for bulk
Bi$_2$Se$_3$.
\section{Bulk Landau Levels}
Liu et al. proposed a 3D Dirac Hamiltonian to describe the bulk states in Bi$_2$Se$_3$~[S1,S2].
This Hamiltonian, written in the basis
$\{\ket{\text{Se}\downarrow},\ket{\text{Bi}\uparrow},\ket{\text{Se}\uparrow},\ket{\text{Bi}\downarrow}\}$, reads:

\begin{equation}\label{eq:fullHkz}
\mathcal{H}=\epsilon(\vec{k})\bf{1}_{4\times 4}+\begin{pmatrix}
  \mathcal{M}(\vec{k}) & \mathcal{A}(k)k_+ & 0 & -\mathcal{B}(k_z)k_z \\
  \mathcal{A}(k)k_- & -\mathcal{M}(\vec{k}) & \mathcal{B}(k_z)k_z & 0 \\
  0  & \mathcal{B}(k_z)k_z & \mathcal{M}(\vec{k}) & \mathcal{A}(k)k_- \\
  -\mathcal{B}(k_z)k_z & 0 & \mathcal{A}(k)k_+ & -\mathcal{M}(\vec{k})
 \end{pmatrix},
\end{equation}
where $\epsilon(\vec{k})=C_0+C_1k_z^2+C_2k^2$, $\mathcal{M}(\vec{k})=M_0+M_1k_z^2+M_2k^2$,
$\mathcal{B}(k_z)=B_0+B_2k_z^2$, $\mathcal{A}(k)=A_0+A_2k^2$, $k_{\pm}=k_x\pm ik_y$ and
$k^2=k_x^2+k_y^2$. In the simplified model, employed in the main text,
we restrict this Hamiltonian by neglecting the $k^3$ terms and
redefine $\Delta\equiv M_0$, $M\equiv M_2$, $C\equiv C_2$, $C_0\equiv 0$ and $v_D=A_0/\hbar$.

To examine the Landau level structure, we introduce the magnetic field (along the $z$ axis) by means of the Peierls
substitution $\vec k\longrightarrow \vec \pi := \vec k+\frac{e\vec A}{\hbar}$, where $\vec A$ is the vector potential,
$\vec B = \nabla \times \vec A$ and $\vec B \| \vec z$. In the quantizing magnetic fields, the LL spectrum of electrons and holes takes
the form of $E_{n,e}(k_z)$ and $E_{n,h}(k_z)$.  The inversion symmetry of the system implies:
$E_{n,e/h}(k_z)=E_{n,e/h}(-k_z)$ and therefore $dE_{n,e/h}/dk_z=0$ at $k_z=0$. This induces a series of singularities in the density
of states for electrons and holes, but importantly for our case, also in the joint density of states (for each LL and inter-LL
resonance, respectively). Using a parabolic expansion in $k_z$ (approximately valid
for $k_z\approx 0$), we get the characteristic $\rho_{e-h}(\omega)\propto1/\sqrt{\hbar\omega-E}$ profiles in the vicinity of each resonance
in the joint density of states. Such a profile is typical of 1D systems with a parabolic dispersion. The singularities in $\rho_{e-h}$,
in reality smoothed by disorder, give rise to the experimentally observed inter-LL resonances. In principle, more complex
Landau level profiles in the momentum $k_z$ may provide further singularities/resonances (due to states at $k_z\neq 0$), nevertheless,
we did not identify any such transitions in our magneto-transmission spectra. The magneto-optical response of Bi$_2$Se$_3$ ($\vec B \| \vec z$)
is thus dominantly determined by $k_z=0$ states.

For $k_z = 0$, the 3D Hamiltonian \eqref{eq:fullHkz} reduces to two 2D Dirac-like Hamiltonians:
\begin{equation}\label{eq:fullH}
\mathcal{H}_0=\begin{pmatrix}
  h_0(\vec{\pi}) & 0 \\
  0 & h_0^*(\vec{\pi}) \end{pmatrix}\,\text{ with }\,
  h_0(\vec{\pi})=\epsilon(\vec{\pi})\bf{1}_{2\times 2}+
  \begin{pmatrix}
   \mathcal{M}(\vec \pi) & A_0\vec \pi _+ \\
   A_0 \vec\pi _- & -\mathcal{M}(\vec\pi)
  \end{pmatrix},
\end{equation}
that will be referred to as the ``full model'' in the following discussion. Now, without $k_z$ terms, we have $\epsilon(\vec\pi)=C\vec\pi ^2$, where $C$ breaks the
particle-hole symmetry. $\mathcal M (\vec \pi)$ is the $k$ dependent mass term $\mathcal M (\vec \pi)=\Delta+M\vec\pi ^2$ with
$\Delta$ determining the band gap $E_g=2\Delta$ at $k=0$.

To study the formation of LLs, we introduce the ladder operators:
\begin{equation}\label{eq:ladder}
a =\frac{l_B}{\sqrt{2}}\vec\pi_-,\,
a^{\dagger} = \frac{l_B}{\sqrt{2}}\vec\pi_+,
\end{equation}
with the magnetic length $l_B=\sqrt{\frac{\hbar}{eB}}$.
These operators obey the standard relations $[a,a^\dagger]=1$, $a\left|n\right\rangle=\sqrt{n}\left|n-1\right\rangle$ and
$a^{\dagger}\left|n\right\rangle=\sqrt{n+1}\left|n+1\right\rangle$. Using the definitions \eqref{eq:ladder},
we can rewrite $h_0(\vec{\pi})$ in terms of the raising and lowering operators:

\begin{equation}\label{eq:h0op}
h_0(a^\dagger,a)=\begin{pmatrix}
                \Delta+\frac{2}{l_B^2}(C+M)(a^\dagger a+\frac{1}{2}) & \frac{\sqrt{2}}{l_B}A_0a^\dagger\\
                \frac{\sqrt{2}}{l_B}A_0a & -\Delta+\frac{2}{l_B^2}(C-M)(a^\dagger a+\frac{1}{2})
               \end{pmatrix}.
\end{equation}
The form of this Hamiltonian suggests the following ansatz for the eigenstates:
\begin{equation}\label{eq:h0states}
 \Phi_{n\neq0} = \begin{pmatrix} c_{n1} \left|n\right\rangle\\ c_{n2}\left|n-1\right\rangle
\end{pmatrix}\,\text{ and }\,
 \Phi_0 = \begin{pmatrix} \left|0\right\rangle\\ 0 \end{pmatrix}\,\text{ with }\, \langle n | m \rangle=\delta_{nm}.
\end{equation}
Solving the Schr\"odinger equation, we find the LL spectrum:
\begin{equation}\label{eq:h0LL}
\begin{split}
 E_{n,\alpha} =&
\frac{M}{l_B^2}+2\frac{C}{l_B^2}n+s_\alpha\sqrt{\left(\frac{C}{l_B^2}+\Delta+2\frac{M}{l_B^2}
n\right)^2+2\frac { A_0^2 }
{l_B^2}n},\\
 E_{0,e} =& \Delta+\frac{C+M}{l_B^2},
\end{split}
\end{equation}
where $s_e=+1$ for electrons and $s_h=-1$ for holes. Note that each state with the energy of $E_{n\geq 1,\alpha}$ is always a
superposition of
the $\ket{\text{Se},\,n,\,\downarrow}$ level and the $\ket{\text{Bi},\,n-1,\,\uparrow}$ level. In contrast, the $E_{0,e}$ state is fully
polarized as a $\ket{\text{Se},\,\downarrow}$ level. The presence of such polarized zero-mode LLs is characteristic of Dirac-type
Hamiltonians and it is well-known, \textit{e.g.}, from physics of graphene~[S3]. However, in graphene, the zero-mode levels
(at $K$ and $K'$ points) are polarized in pseudospin not in real spin as in the case of Bi$_2$Se$_3$.

The Landau levels for the $h_0^*$ Hamiltonian are found in an analogous way. Here, we take
\begin{equation}\label{eq:h0*states}
 \Phi_{n\neq0}^* = \begin{pmatrix} c_{n1*} \left|n-1\right\rangle\\ c_{n2*}\left|n\right\rangle
\end{pmatrix}\,\text{ and }\,
 \Phi_0^* = \begin{pmatrix} 0\\ \left|0\right\rangle \end{pmatrix}\,\text{ with }\, \langle n | m \rangle=\delta_{nm}
\end{equation}
as the ansatz for the eigenstates and get LLs for the $h_0^*$ Hamiltonian:
\begin{equation}\label{eq:h0*LL}
\begin{split}
 E_{n,\alpha}^* =&
-\frac{M}{l_B^2}+2\frac{C}{l_B^2}n+s_\alpha\sqrt{\left(\frac{C}{l_B^2}-\Delta-2\frac{M}{l_B^2}
n\right)^2+2\frac { A_0^2
}
{l_B^2}n},\\
 E_{0,h}^* =& -\Delta+\frac{C-M}{l_B^2}.
\end{split}
\end{equation}

Note that ``$^*$'' just denotes energies/states belonging to the $h_0^*$ Hamiltonian and does not stand for complex conjugation. We will keep this notation
in the next sections. For the $h_0^*$ Hamiltonian, the state with the energy $E_{n\geq 1,\alpha}$ is always a superposition of the
$\ket{\text{Se},\,n-1,\,\uparrow}$ level and the $\ket{\text{Bi},\,n,\,\downarrow}$ level. Again, the zero-mode LL is fully
spin-polarized, in this case as a $\ket{\text{Bi},\,\downarrow}$ state.

\begin{figure}[t]
\includegraphics[width=0.55\textwidth]{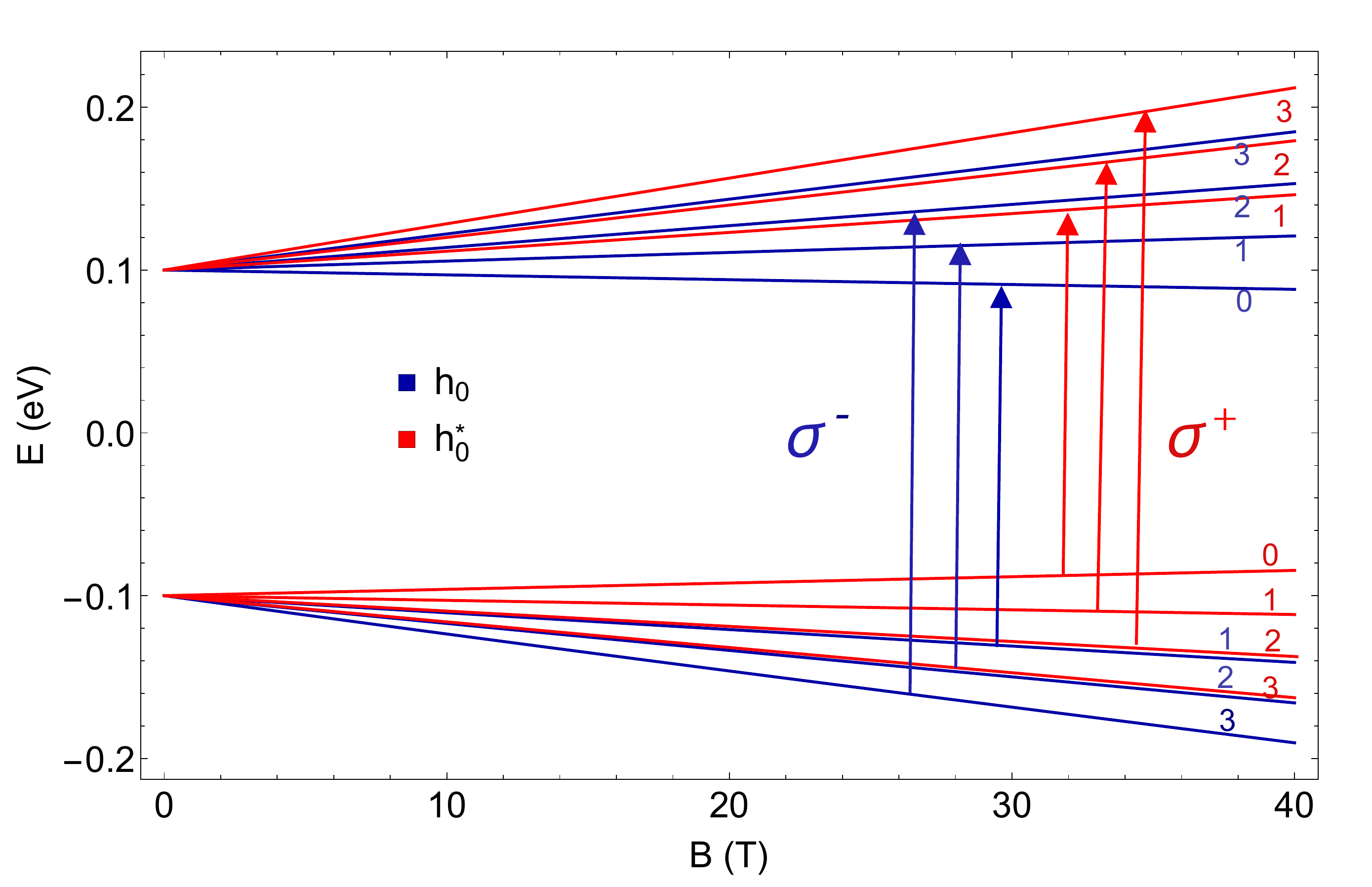}
\caption{The Landau level spectrum of the full model \eqref{eq:fullH}. Blue levels originate from the
$h_0$ sub-Hamiltonian, red levels from the $h_0^*$ Hamiltonian. The arrows show the dominant type of electric-dipole
 transitions for the respective Hamiltonian. \label{pic:LLfull}}
\end{figure}

The LL spectrum of the Hamiltonian \eqref{eq:fullH} (calculated within the full model) is plotted in Fig.~\ref{pic:LLfull} for
parameters derived from our magneto-optical experiments, together with electric-dipole transitions discussed later on in detail.

\begin{figure}[htb]
\subfigure{\includegraphics[width=0.49\textwidth]{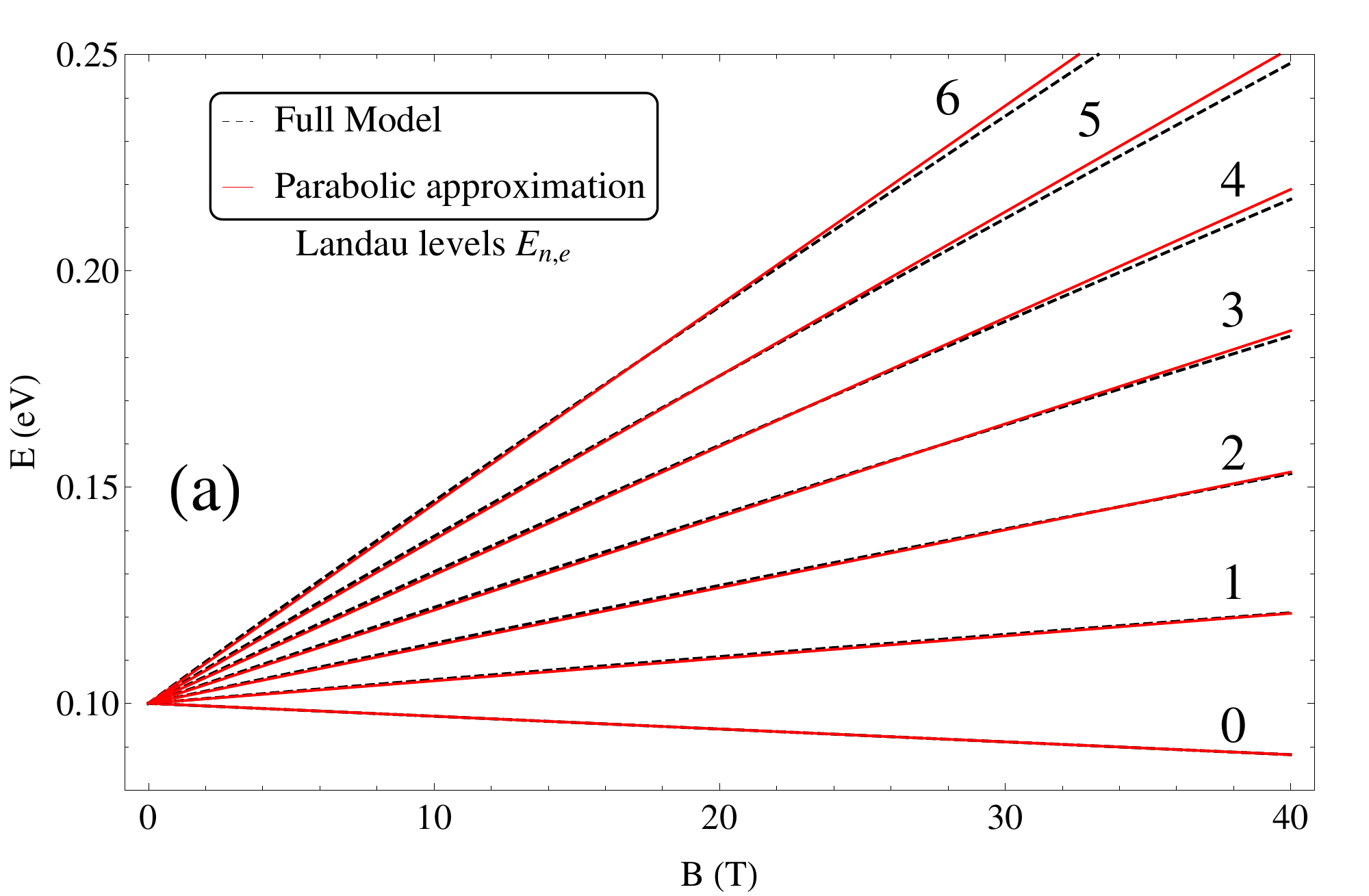}}\hfill
\subfigure{\includegraphics[width=0.49\textwidth]{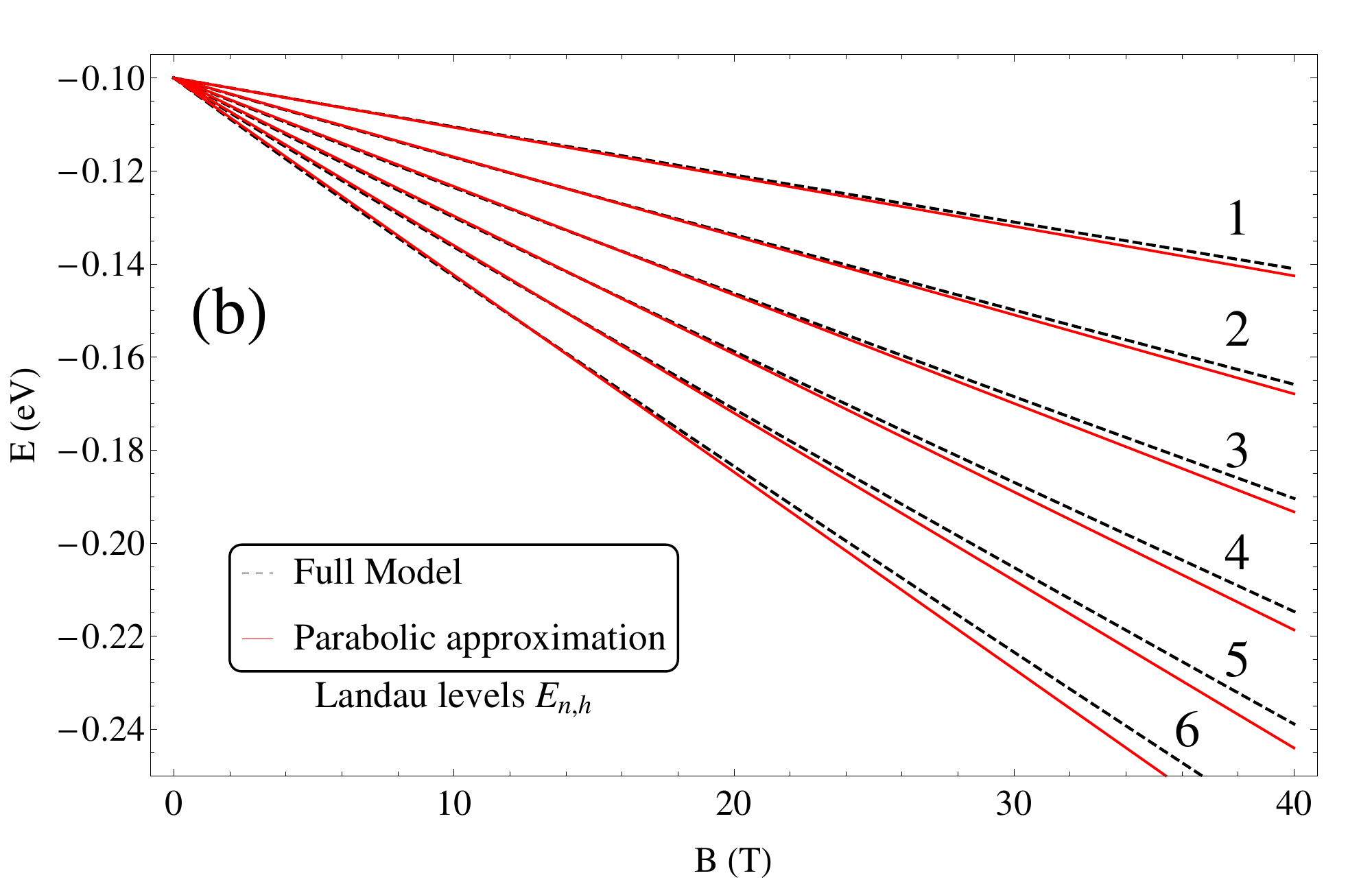}}
\caption{LLs calculated within the full model, Eq.~\eqref{eq:h0LL} and \eqref{eq:h0*LL}, compared to LLs obtained
by expansion of the same formulas for small magnetic field (\textit{i.e.}, by linearization in $B$), see Eq.~\eqref{eq:LLapprox}.
Significant deviations appear only at high magnetic fields and high LL indices. For simplicity,
only the levels of the $h_0$ Hamiltonian are shown.\label{pic:LLapproximation}}
\end{figure}

To simplify the LL spectra of the $h_0$ and $h_0^*$ Hamiltonians, we assume perfectly parabolic bands ($A_0^2=\hbar^2v_D^2=-4 M \Delta$)
and expand Eq.~\eqref{eq:h0LL} and Eq.~\eqref{eq:h0*LL} for small magnetic fields. This way we get LLs strictly linear
in the applied magnetic field:
\begin{equation}\label{eq:LLapprox}
\begin{split}
 E_{n,e} =& E_{0,e}+\hbar\omega_c^en, \\
 E_{n+1,e}^* =& E_{n,e}+\hbar\omega_c^D, \\
 E_{n+1,h} =& E_{n,h}^*-\hbar\omega_c^D,\\
 E_{n,h}^* =& E_{0,h}^*-\hbar\omega_c^hn,
\end{split}
\end{equation}
where the cyclotron frequencies are defined as $\omega_c^{e/h}=eB/m_{e/h}\,$ and $\,\omega_c^D=eB/m_D$ with the effective
masses $m_{e/h}=2\hbar^2/\left(\hbar^2/m_D\pm4C\right)$ and the Dirac mass $m_D=\Delta/v_D^2$, respectively. Importantly, for parameters
deduced from our experimental data, this simplified
LL spectrum is nearly identical to that calculated within the full model, see Fig.~\ref{pic:LLapproximation}.

\begin{figure}[htb]
\subfigure{\includegraphics[width=0.49\textwidth]{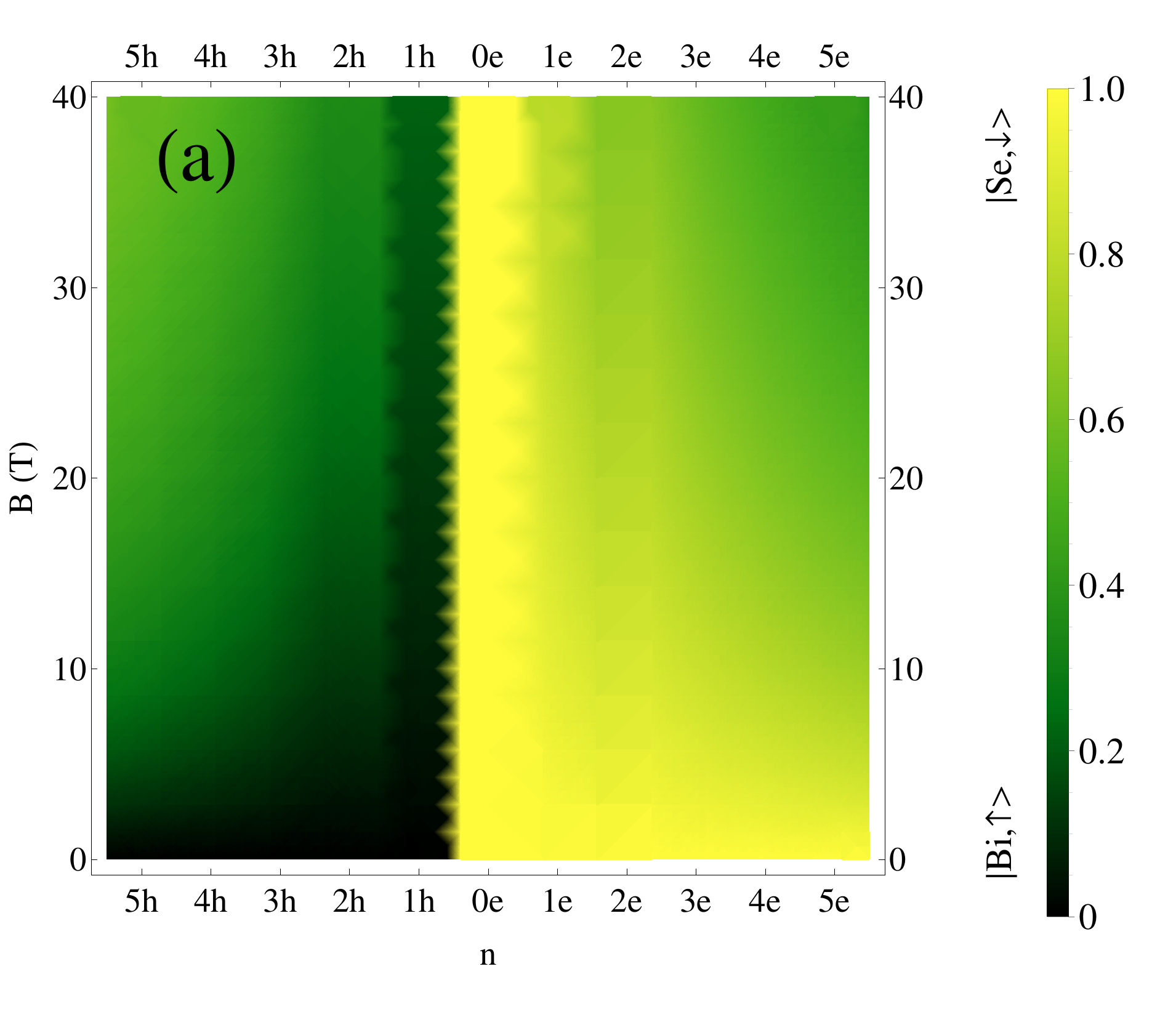}}\hfill
\subfigure{\includegraphics[width=0.49\textwidth]{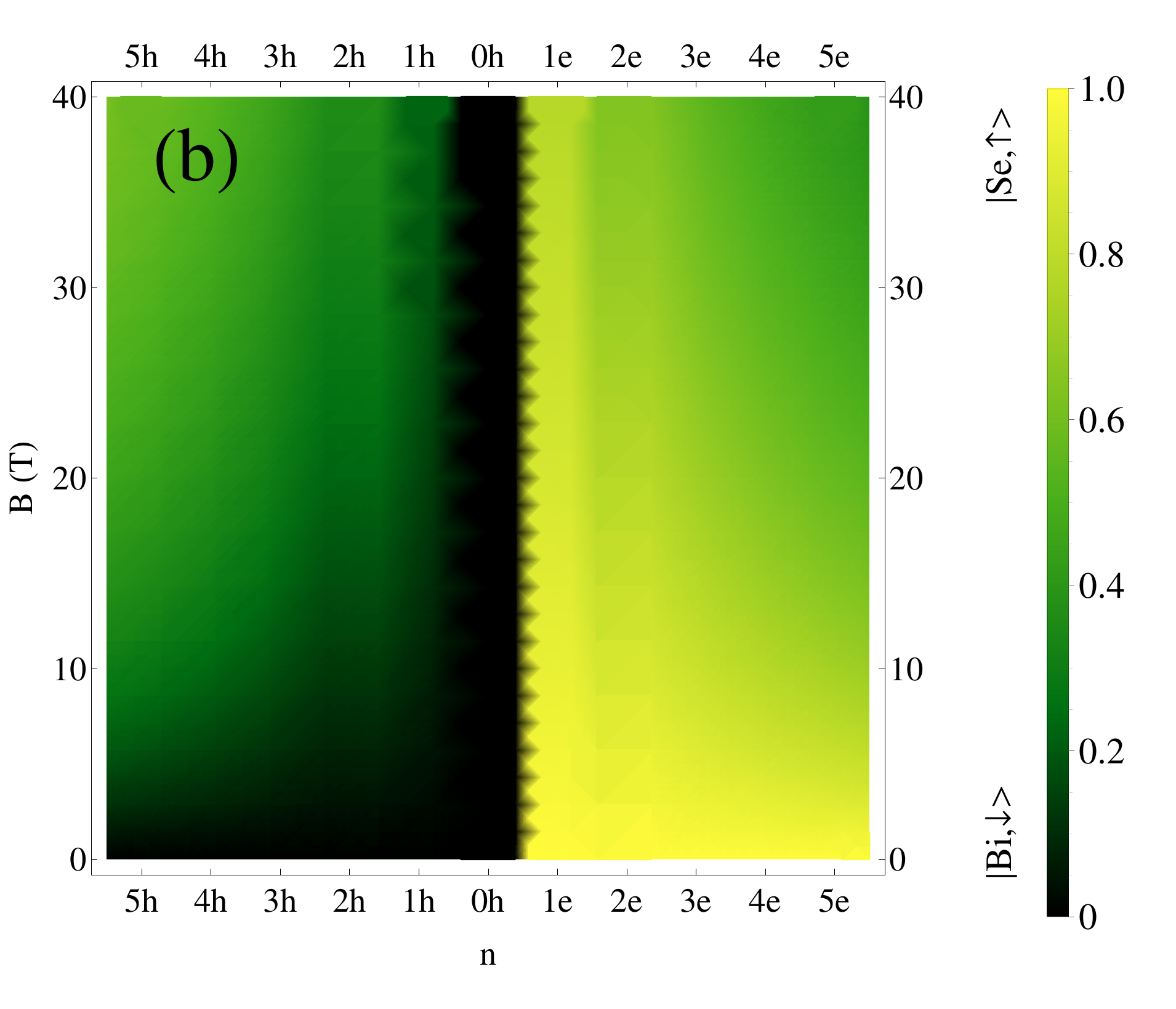}}
\caption{Probability density of the energy states being Bi-like (dark color) or Se-like (light color) with corresponding
spin-polarization. $ne$ and $nh$
denote the LL with index $n$ for electron ($e$) or hole ($h$) states. This probability is shown in the part (a) for LLs belonging to the $h_0$
Hamiltonian, where these levels are a mixture of $\ket{\text{Se},\,\downarrow}$ and $\ket{\text{Bi},\,\uparrow}$ states. The part
(b) shows this probability for the $h_0^*$ Hamiltonian, where LLs are a mixture of
$\ket{\text{Se},\,\uparrow}$ and $\ket{\text{Bi},\,\downarrow}$ states.
Apart from the zero-mode LLs, one can see that the spin-polarization becomes weaker with increasing
magnetic field. However for magnetic fields $B\lesssim 40\,$T and low level indices the dominant spin
polarization of the energy states stays the same  as in the $B\longrightarrow0$\,T limit.\label{pic:densities}}
\end{figure}

Further information about Landau levels may be obtained from the analysis of individual eigenstates,
which (with the exception of zero-mode levels) represent a superposition of spin-up and spin-down states
in the conduction and valence bands. Nevertheless, for low-energy/low index levels we may always find
the dominant state in this superposition, see Fig.~\ref{pic:densities}. We may conclude that
the eigenstates of the $h_0$ Hamiltonian with the energies of $E_{n,e}$ and $E_{n,h}$ can be considered
as selenium-like spin-down ($E_n^{\text{Se}\downarrow}$) and bismuth-like spin-up ($E_{n-1}^{\text{Bi}\uparrow}$)
levels, respectively. Analogously, states with the energies of $E_{n,e}^*$ and $E_{n,h}^*$ have
Se-like spin-up ($E_{n-1}^{\text{Se}\uparrow}$) and bismuth-like spin-down ($E_{n}^{\text{Bi}\downarrow}$) character, respectively.
This assignment of spin-projections will facilitate the definition of $g$ factors in the next section.

\section{Definition of g factors}

For $B=0$\,T the energy states in our system are spin degenerate as required by time- and inversion- symmetry of Bi$_2$Se$_3$.
When the magnetic field is applied, this spin degeneracy of states is lifted due
to the magnetic moment $\vec \mu$ of electrons. This splitting may be described by a corresponding effective $g$ factor in
the Zeeman term, $E_Z=-\vec\mu\vec B$, with $\vec\mu=g\mu_B\vec s/\hbar$. The total $g$ factor comprises three contributions. As shown in our
experiments, and by the subsequent data analysis, the main contribution results from the strong the spin-orbit coupling in Bi$_2$Se$_3$,
which is inherently included within the Hamiltonian \eqref{eq:fullHkz}. Further (minor)
corrections come from the free electron $g_0\approx 2$ factor (free-electron Zeeman term)
and a perturbative contribution from remote energy bands~[S4].

From the previous section, we know that, for small level indices $n$ and low energies, LLs are nearly spin polarized.
In addition, the Landau levels in the valence and conduction bands are bismuth- and selenium-like, respectively. Following this fact,
we can express the $g$ factors of charge carriers, in the conduction (c) and valence (v) bands, in terms of LLs calculated from the Hamiltonian Eq.~\eqref{eq:fullH}:
\begin{subequations}\label{eq:gfactors}
\begin{alignat}{2}
\label{eq:ge}
g_c=g_e=& g_\text{Se}(n,B)=(E_n^{\text{Se}\uparrow}-E_n^{\text{Se}\downarrow})/(\mu_BB)=(E_{n+1,e}^{*}-E_{n,e})/(\mu_B B),\\ \label{eq:gh}
g_v=-g_h=& g_\text{Bi}(n,B)=(E_n^{\text{Bi}\uparrow}-E_n^{\text{Bi}\downarrow})/(\mu_BB)=(E_{n+1,h}-E_{n,h}^{*})/(\mu_B B).
\end{alignat}
\end{subequations}
This definition of the $g_{e/h}$ factors for electrons and holes is consistent with magnetic-dipole selection rules: $n\rightarrow n\pm1$,
which interconnect states belonging to the $h_0$ and $h_0^*$ Hamiltonians~[S5].
In addition, we may crosscheck this definition by taking genuine Dirac particles (electrons) in the vacuum. For this, we have to take $A_0=\hbar c$ ($v_D=c$),
$\Delta=m_0 c^2$ and $M=C=0$, where $c$ is the speed of light and $m_0$ the rest mass of a free electron. Indeed, we get $g_0=2$ as expected.

The definition~\eqref{eq:gfactors} implies $g$ factors that, in general, depend on the magnetic field as well as on the LL index. However,
within our parabolic approximation ($\hbar^2v_D^2=-4M\Delta$) and for LLs linearized in $B$, see Eq.~\eqref{eq:LLapprox}, the spin-splitting of electrons and holes
becomes linear in magnetic field, $E_s=\hbar\omega_c^D$, implying thus $g_c=-g_v=g_e=g_h=2m_0/m_D=2m_0 v_D^2/\Delta$. For the experimentally determined Dirac mass $m_D=(0.080\pm0.005)m_0$
we get $g_e=g_h\approx25$.

Assuming the parabolic approximation ($\hbar^2v_D^2=-4M\Delta$), but taking the full expressions for Landau levels \eqref{eq:h0LL} and \eqref{eq:h0*LL}, \textit{i.e.}, not linearized
in $B$, the $g$ factors for electrons and holes slightly differ and also gain a weak magnetic-field
and Landau-level-index dependence, see Fig.~\ref{pic:gfactors}. Nevertheless, this is not sufficient to account for the experimentally observed
difference, $g_e-g_h\approx3$, derived from our data in the main text. This may only be explained by further corrections (the Zeeman term with the free-electron $g_0=2$ factor and the
influence of remote bands).

\begin{figure}[htb]
\subfigure{\includegraphics[width=0.49\textwidth]{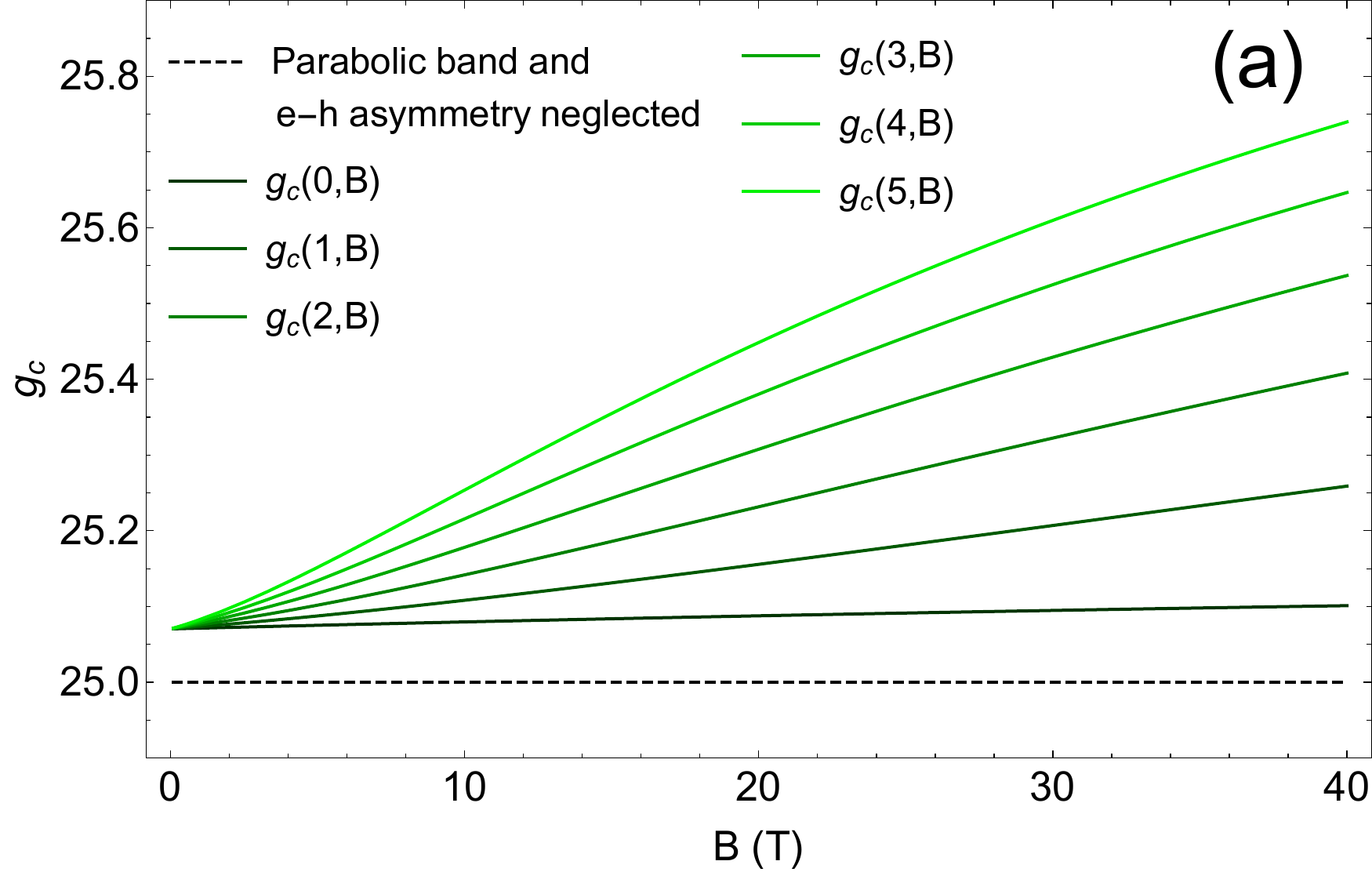}}\hfill
\subfigure{\includegraphics[width=0.49\textwidth]{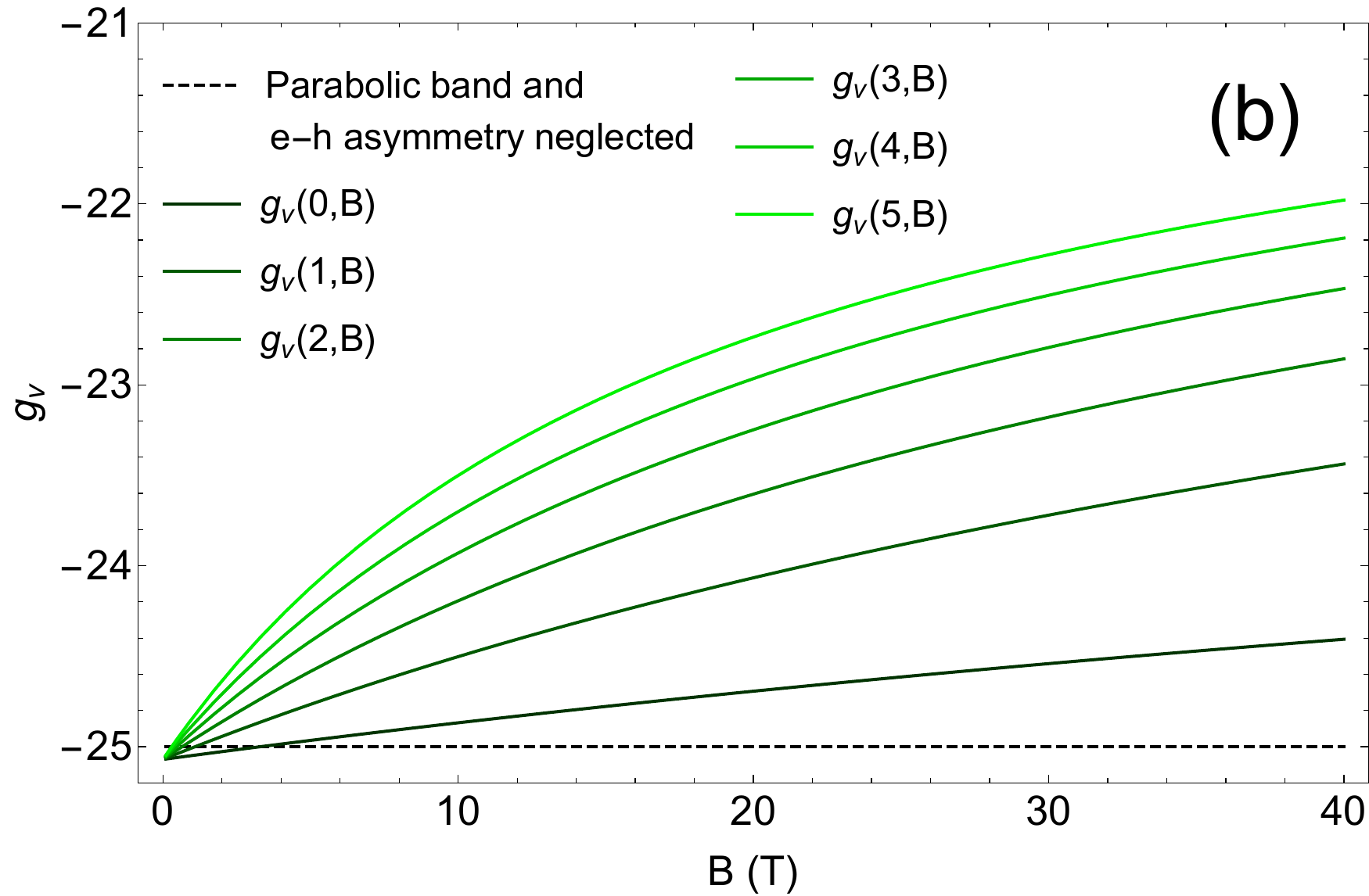}}
\caption{$g$ factor for charge carriers in the conduction  and valence bands, $g_c(n,B)$ and  $g_v(n,B)$, exhibiting a weak dependence on the level index $n$
and the magnetic field strength $B$ according to equations \eqref{eq:ge} and \eqref{eq:gh}. (a) With $n\leq 5$
and $B\leq 40$\,T the approximate constant value for $g_c$ shows a deviation of $\lesssim\,3\,\%$ from the $n$-
and $B$-dependent $g$ factor. (b) For $g_v$ this deviation grows to $\lesssim\,12\,\%$.
Let us note that we use the notation $g_c=g_e$ and $g_v=-g_h$.
\label{pic:gfactors}}
\end{figure}

\section{Zero-mode Landau levels in high magnetic fields}

In high magnetic fields, the spin-polarized zero-mode LLs $E_{e,0}=\Delta+(C+M)/l_B^2$ and
$E_{h,0}=-\Delta+(C-M)/l_B^2$ approach each other and become well separated from the rest of levels.
In such a case, we can  describe them by an effective Hamiltonian:
\begin{equation}\label{eq:h00}
\mathcal{H}_{\mathrm{zero-mode}}=\begin{pmatrix}
                E_{e,0} & -\hbar\tilde{v}_D k_z\\
                -\hbar\tilde{v}_D k_z & E_{h,0}
               \end{pmatrix},
\end{equation}
which can be derived from the ($k_z$ dependent) 3D Dirac Hamiltonian~\eqref{eq:fullHkz}, proposed in Refs.~[S1,S2],
in which we neglect the terms cubic in $k$, terms square in $k_z$ and introduce the magnetic field via Peierls substitution.
Taking account of the experimentally observed uniaxial anisotropy of Bi$_2$Se$_3$~[S6,S7], one can assume
that $\tilde{v}_D=B_0/\hbar\lesssim v_D$. The effective Hamiltonian~\eqref{eq:h00} is equivalent
to a 1D Dirac-type Hamiltonian with the band gap of $E_{e,0}-E_{h,0}$. This gap vanishes at the crossing field
$B_c=\hbar \Delta/|eM|\approx300$~T, when the system changes from a (narrow gap) semiconductor into a gapless
semimetal. The electronic bands in such a semimetal are equivalent to the 1D Dirac-type channel, $E(k_z)=\pm \hbar\tilde{v}_D|k_z|$
with a strong (LL) degeneracy of states $eB/h$.

\section{Selection rules and matrix elements}

To describe the response of our system to an externally applied electromagnetic field, we employ
the standard (linear-response) Kubo-Greenwood formalism. For $\sigma^\pm$ polarized radiation,
the optical conductivity tensor, in a system with eigenstates $\ket{\Psi_n}$ and corresponding energies $E_n$, reads:
\begin{equation}\label{eq:kubo}
\sigma_{\pm}(\omega,B)\propto i\frac{B}{\omega}\sum_{n,n'}\left(\left(f_n-f_{n'}\right)\frac{\betrag{\langle\Psi_{n'}|\hat{v}_\pm(B)|\Psi_n\rangle}^2}{E_n-E_{n'}-\hbar\omega+i\gamma}\right),
\end{equation}
where $f_n$ is the occupation factor, $\gamma$ the phenomenological broadening parameter and $\hat{v}_\pm$
the velocity operators. The matrix elements $\langle\Psi_{n'}|\hat{v}_\pm(B)|\Psi_n\rangle$ determine
the active electric-dipole transitions (selection rules) between different eigenstates.

The velocity operator can be directly obtained from Eq.~\eqref{eq:fullH} by calculating
$\hat v_i=\frac{1}{\hbar}\frac{\partial H_0}{\partial \pi_i }$, where $i=x,y$ and $v_\pm=(v_x\pm i v_y)/\sqrt{2}$:
\begin{equation}
\hat{v}_+=\begin{pmatrix}
 \frac{2(C+M)}{\hbar l_B}a^\dagger & 0 & 0 & 0 \\
 \sqrt{2}\,v_D & \frac{2(C-M)}{\hbar l_B}a^\dagger & 0 & 0\\
 0 & 0 & \frac{2(C+M)}{\hbar l_B}a^\dagger & \sqrt{2}\,v_D \\
 0 & 0 & 0 & \frac{2(C-M)}{\hbar l_B}a^\dagger
\end{pmatrix}\,=\,\hat v_-^\dagger.
\end{equation}
Notably, these velocity operators (with two independent $2\times2$ diagonal blocks) imply that the electric-dipole transitions
are not active between pairs of LLs belonging to different Hamiltonians $h_0$ and $h_0^*$. This is different
from magnetic-dipole transitions, which connect states originating in different Hamiltonians $h_0$ and $h_0^*$.

Taking the eigenstates $\ket{\Psi_n}$, \textit{i.e.}, LLs expressed by Eqs.~\eqref{eq:h0states} and \eqref{eq:h0*states}
arranged as:
\begin{equation}
 \ket{\Psi_n}=\begin{pmatrix}
              \Phi_n \\
              \Phi_n^*
             \end{pmatrix},
\end{equation}
we may calculate the selection rules sensitive to the circular polarization of the infrared radiation:
\begin{equation}
\begin{split}
\bra{\Psi_{n'}}\hat{v}_+(B)\ket{\Psi_n}&=\mathcal{F}_n\delta_{n,n'-1}+\mathcal{F}_n^*\delta_{n,n'-1}\,\text{ with }\,n\geq 0, \\
\bra{\Psi_{n'}}\hat{v}_-(B)\ket{\Psi_n}&=\mathcal{F}_n\delta_{n,n'+1}+\mathcal{F}_n^*\delta_{n,n'+1}\,\text{ with }\,n\geq 1,
\end{split}
\end{equation}
where $\mathcal{F}_n=\mathcal{F}_n(A_0,\Delta,M,C,B)$ and $\mathcal{F}^*_n=\mathcal{F}^*_n(A_0,\Delta,M,C,B)$ are amplitudes
belonging to transitions within the $h_0$ and $h_0^*$ Hamiltonians, respectively.  Importantly, we get the same selection rules for electric-dipole
transitions between LLs belonging to the $h_0$ and $h_0^*$ Hamiltonians, $n\rightarrow n\pm1$. Nevertheless, the corresponding amplitudes
$\mathcal{F}_n$ and $\mathcal{F}^*_n$ may strongly differ. Here we should again recall that the ``$^*$'' symbol
refers to the given sub-Hamiltonian $h_0^*$ and does not denote the complex conjugation ($|\mathcal{F}_n^*|\neq|\mathcal{F}_n|$).
The difference in amplitudes is clearly seen in the matrix elements for interband inter-LL
absorption, see Fig.~\ref{pic:matel}. The interband inter-LL absorption between levels belonging to the Hamiltonians
$h_0$ and $h_0^*$ is dominantly active in $\sigma^-$ and $\sigma^+$ polarized radiation, respectively.

\begin{figure}[htb]
\includegraphics[width=0.55\textwidth]{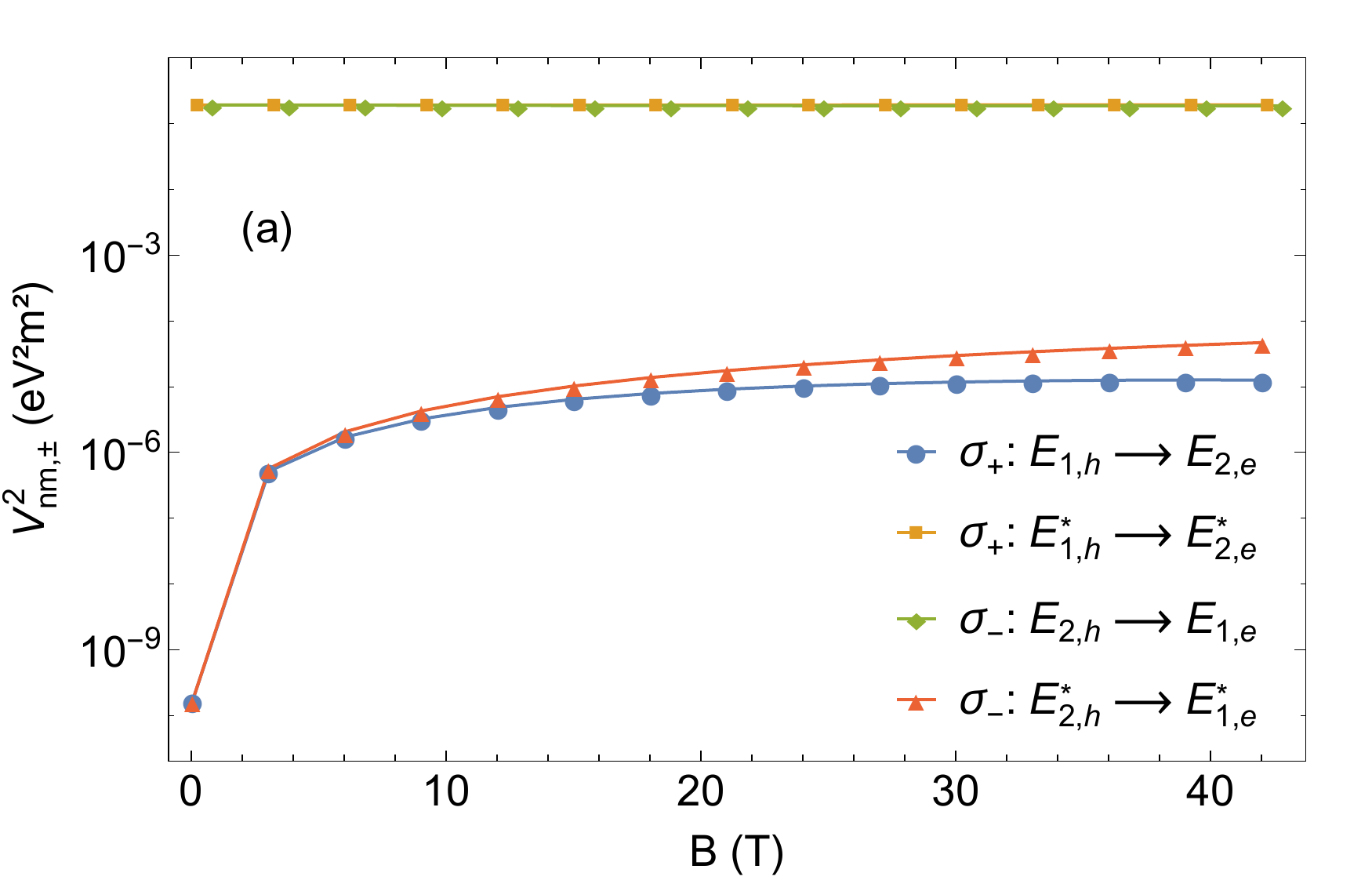}
\caption{Matrix elements $V_{nm,\pm}^2=\hbar^2\left|\bra{\Psi_m}\hat{v}_\pm(B)\ket{\Psi_n}\right|^2$ for interband inter-LL
transitions in Bi$_2$Se$_3$ plotted in the logarithmic scale (for $m,n=1,2$). 
For $\sigma^-$ and $\sigma^+$ polarized light, interband absorption is dominated by transition between LLs originating from the $h_0$ and $h_0^*$ sub-Hamiltonian, respectively.
\label{pic:matel}}
\end{figure}

\section{Energy band gap of Bi$_2$Se$_3$}

The magneto-transmission experiment, presented in this paper, provides a fairly precise estimate of the energy band gap in Bi$_2$Se$_3$: $2\Delta=(200\pm4)$~meV.
This result is in very good agreement with other optical studies. For instance, it matches well the value of $\sim175$ and 160~meV expected for the band gap at low temperatures,
as extracted from extensive reflectivity measurements on a series of bulk specimens with different electron densities in Refs.~[S8] and [S9], respectively. Similarly, our results correspond very well to conclusions of recent low-temperature infrared transmission studies performed on thin layers of Bi$_2$Se$_3$ (a series of samples with thicknesses below 100~nm) prepared by molecular beam epitaxy on a (111) oriented silicon substrate~[S10].

On the other hand, ARPES studies of bulk Bi$_2$Se$_3$ report gap values close to 300~meV, see, \textit{e.g.}, Refs.[S11-13],
which are significantly higher as compared to our results and other optical studies, which may invoke questions about the nature of the thin Bi$_2$Se$_3$ samples studied
in these works. It is, for instance, not a priori clear whether and how the substrate properties and particular growth conditions influence the observed energy band gap and the overall band structure.

\begin{figure}[t]
\includegraphics[trim = 25mm 180mm 45mm 20mm, clip=true, width=10cm]{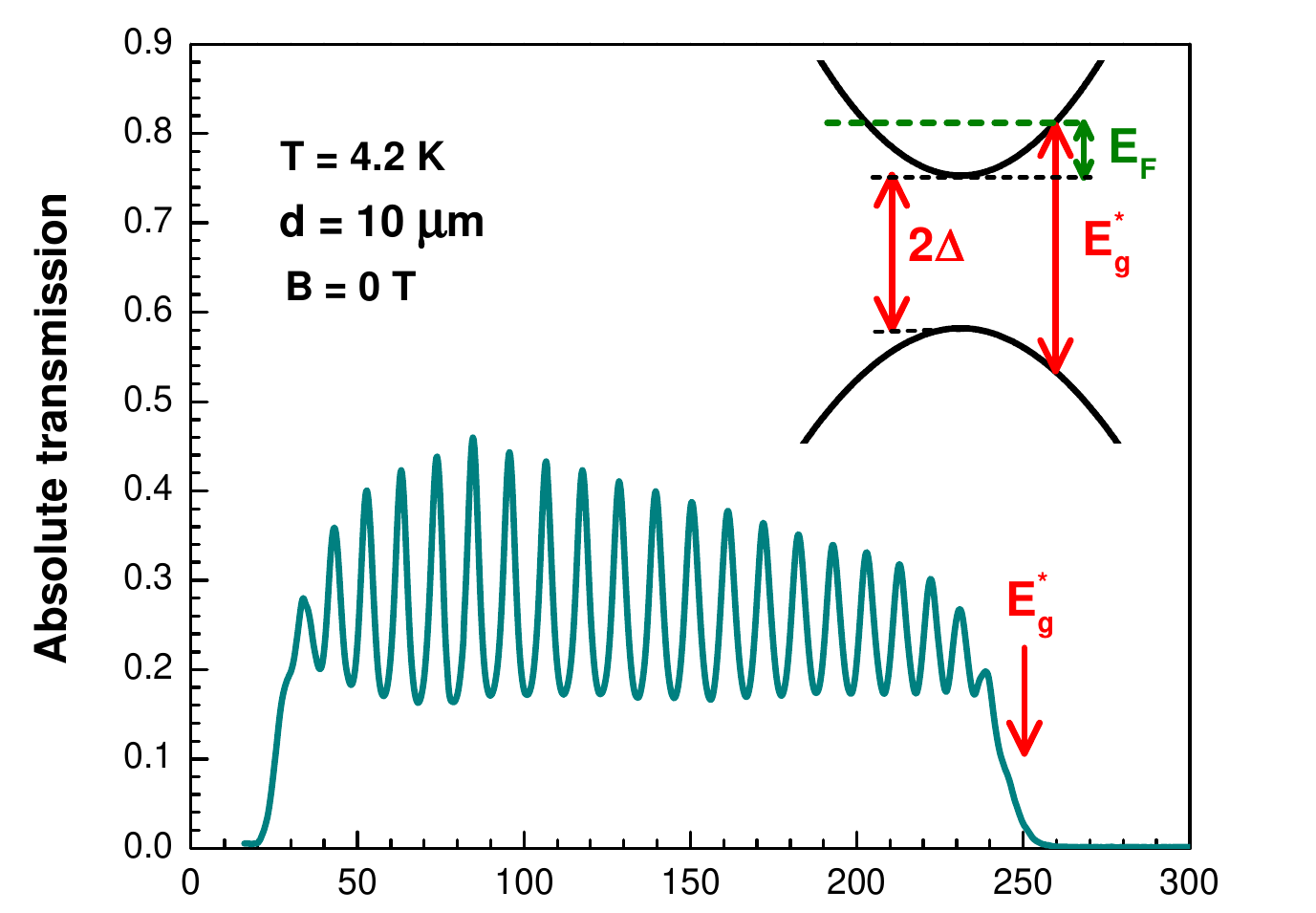}
\caption{Low temperature infrared transmission spectrum of a 10-micron-thick free-standing layer of Bi$_2$Se$_3$. The inset schematically shows the relation between
the high energy cut-off of the transmission spectrum and the electronic band gap:  $E_g^*>2\Delta$. The pronounced modulation of the spectrum corresponds to the Fabry-P\'{e}rot
oscillations, showing rather high crystalline quality of the studied Bi$_2$Se$_3$ bulk specimen.\label{TR}}
\end{figure}

To provide another independent verification of the band gap value deduced optically, we have performed low-temperature infrared transmission measurements on
thin self-standing Bi$_2$Se$_3$ layers, prepared simply by slicing bulk crystals. A typical infrared transmission spectrum is plotted in Fig.~\ref{TR}.
This spectrum has been measured on a 10~$\mu$m-thick specimen, with the electron concentration close to 10$^{18}$~cm$^{-3}$. This density has been deduced
from magneto-transport measurements and it is comparable (slightly higher) with respect to the electron density in the thin film of Bi$_2$Se$_3$ studied in our magneto-transmission
experiments.

The presented transmission spectrum exhibits a fairly sharp high energy cut-off at $E_g^\star \approx 250$~meV. This provides us with a well-defined \textit{upper bound}
for the electronic band gap of Bi$_2$Se$_3$, $2\Delta<E_g^\star$, as schematically depicted in the inset of Fig.~\ref{TR}. Clearly, this high-energy cut-off is significantly below the band gap of 300 meV, which is deduced from ARPES measurements. Instead, in our case, the band gap should approach $2\Delta\approx E_g^\star-2E_F\approx200-210$ meV, as implied by the Burstein-Moss shift in materials with high electron-hole symmetry ($m_e\sim m_h$). The Fermi level has been estimated as $E_F=20-25$~meV for the given electron density.

To conclude, the ARPES data indicate the band gap, which is significantly higher as compared to rather direct optical measurements presented in this paper as well as those performed by other groups. At present, we do not have a clear explanation for this intriguing difference, nevertheless, we speculate that ARPES is, as a matter of fact, a surface-sensitive technique. As such, the deduced band gap might be influenced by specific band-bending effects on the samples' surfaces, notably in the system with an inversed order of electronic bands. This difference clearly shows that the consensus about the size of the band gap in Bi$_2$Se$_3$ has not yet been established. This includes also on-going discussions, one versus another ARPES data, about the direct/indirect nature of the band gap in this material, see Refs.~[S11-15].

\begin{figure}[t]
\includegraphics[trim = 05mm 18mm 25mm 20mm, clip=true, width=12cm]{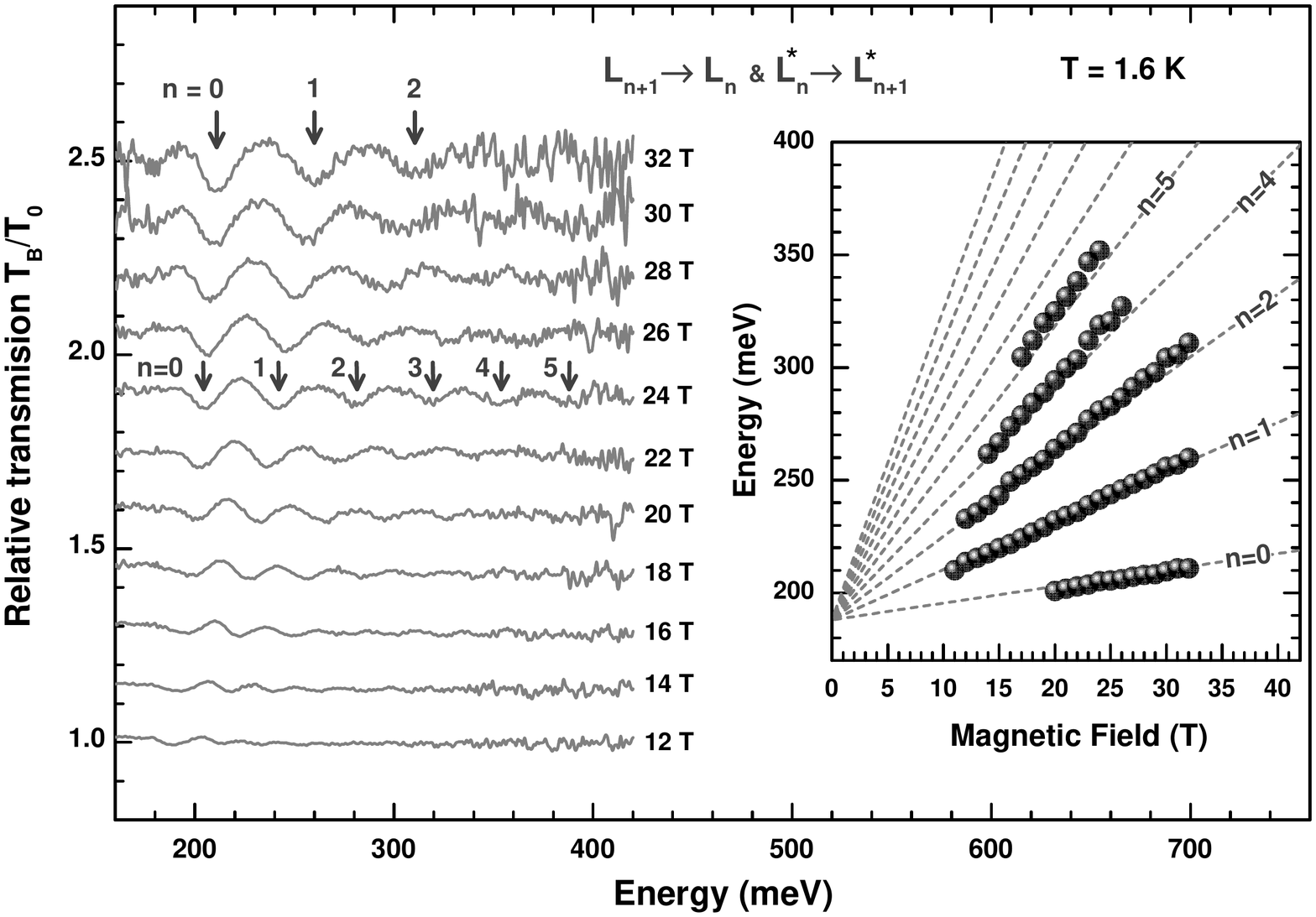}
\caption{Relative transmission spectra of Bi$_2$Se$_3$ in the middle infrared spectral range plotted for selected values of the magnetic field.
      At $B=24$ and 32~T, individual inter-LL excitations are denoted by vertical arrows and identified by the corresponding index $n$. The fanchart of the observed resonances
      is plotted in the inset. The dashed lines represent theoretical fits described in the text (for $C\equiv0$).\label{Interband410}}
\end{figure}

\section{Additional experimental data}

Here we present complementary experimental data, obtained in high-field infrared magneto-transmission experiments performed on a 102~nm-thick Bi$_2$Se$_3$ layer on a InP(111)B substrate.
This sample was prepared using MBE technique under conditions analogous to the 290-nm-thick sample described in the main part of the paper and it is weakly $n$-doped with the electron density slightly
below $10^{18}$~cm$^{-3}$. In spite of a lower signal-to-noise ratio obtained on this thinner sample, the observed magneto-optical response, see Figs.~\ref{Interband410} and \ref{Intraband410}, allows us to draw the same conclusions about the electronic band structure of Bi$_2$Se$_3$ as in the case of the 290-nm-thick specimen.

\begin{figure}[t]
\includegraphics[width=8cm]{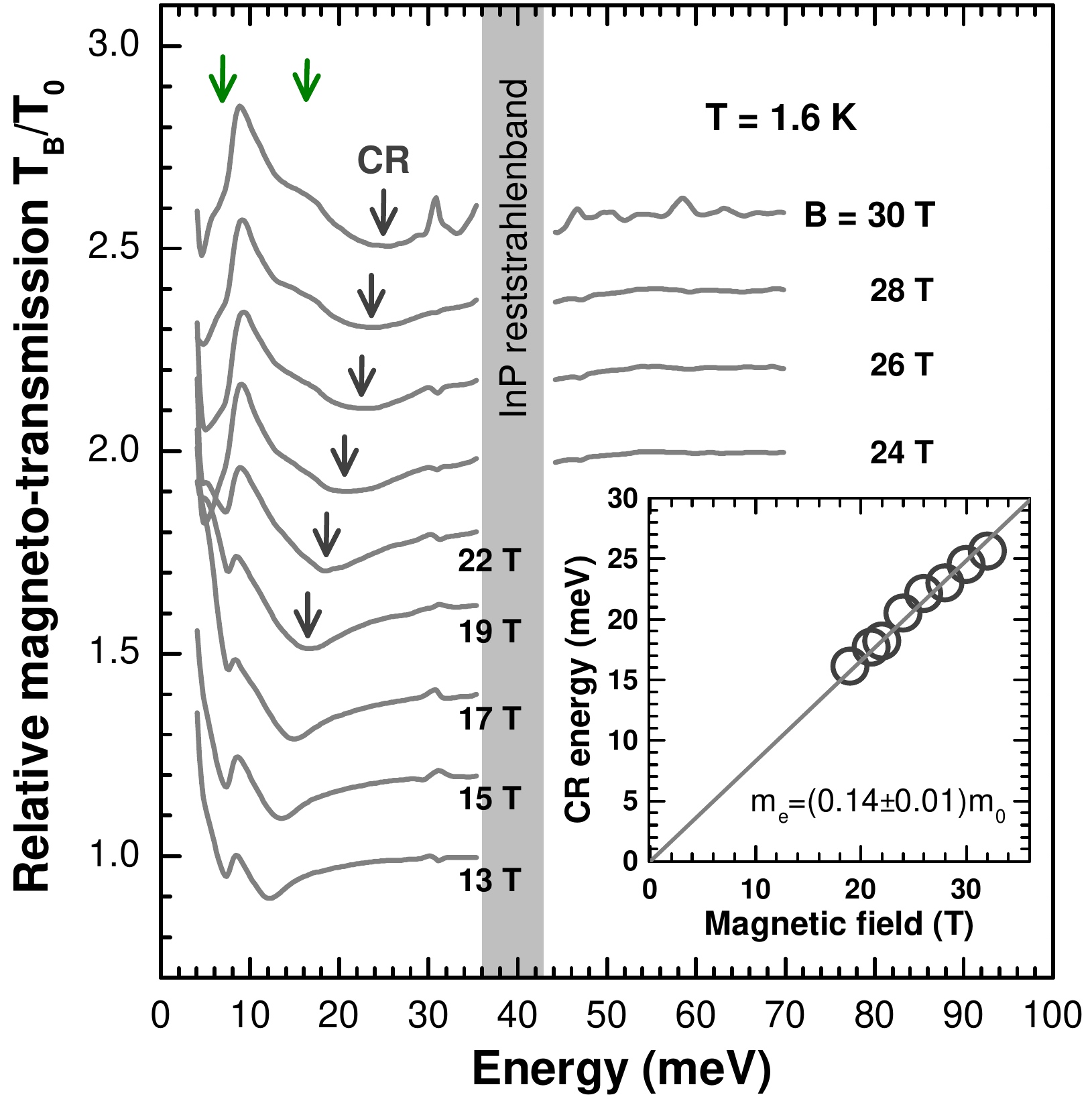}
\caption{Far infrared magneto-transmission spectra of the 102-nm thick Bi$_2$Se$_3$ on a InP substrate. The CR absorption is manifested
as a dip in the relative magneto-transmission spectra $T_B/T_0$ (denoted by vertical arrows) and follows linear in $B$ dependence, which implies
the effective mass of electrons $m_e=(0.14\pm0.01)m_0$, see the inset. At higher magnetic fields, the response at low energies
is characterized by field-induced transmission, $T_B/T_0>1$, which is due to suppressed zero-field Drude-type absorption, see Ref.~[16]
for analogous behavior in highly doped graphene. At low energies, a pronounced effect of interaction between CR
and $\alpha$ and $\beta$ phonon modes are also observed (cf. Fig.~1 in the main part of the paper).\label{Intraband410}}
\end{figure}

The overall linear in $B$ optical response, including intraband and interband inter-LL resonances, points towards parabolic profiles
of both conduction and valence bands. The Dirac mass and the band gap (derived from the separation and low-magnetic-field extrapolation of resonances in Fig.~\ref{Interband410}, respectively)
as well as the effective mass of electrons (read from CR absorption in Fig.~\ref{Intraband410}) are nearly identical to values obtained from the 290-nm-thick sample: $m_D=(0.080\pm0.005)m_0$,
$2\Delta=190\pm5$~meV and $m_e=(0.14\pm0.1)m_0$. The parabolic profiles of electronic bands together with the condition $m_e \approx 2m_D$ thus imply also for this Bi$_2$Se$_3$ sample
the specific match between spin-splitting and cyclotron energy ($E_s=2E_c$).

In our deeper analysis, we compared (fitted) the experimentally read positions of resonances with theoretically expected transition energies calculated using full (non-linearized) expressions for LLs Eqs.~\eqref{eq:h0LL} and \eqref{eq:h0*LL}, see the inset of Fig.~\ref{Interband410}. The best agreement was found for parameters $v_D=(0.45\pm0.03)\times10^6$~m.s$^{-1}$, $\Delta=(0.095\pm0.003)$~eV and $M=-(22.5\pm1.5)$~eV.\AA$^2$, which practically match those deduced from the 290-nm-thick sample, see the main text. The deviation from the condition $\hbar^2 v_D^2=-4M\Delta$ for the exact parabolicity of electronic bands does not exceed a few percent for this set of parameters. This provides us with another justification for
our approximation in which we describe the band structure of Bi$_2$Se$_3$ using a simplified Hamiltonian implying only two parameters: the band gap $2\Delta$ and the velocity parameter $v_D$.

\vspace{1cm}

\textbf{References:}\\

[S1] H. Zhang \textit{et al.}, Nature Phys. \textbf{5}, 438 (2009).\newline
[S2] C.-X. Liu \textit{et al.}, Phys. Rev. B \textbf{82}, 045122 (2010).\newline
[S3] M. O. Goerbig, Rev. Mod. Phys. \textbf{83}, 1193 (2011).\newline
[S4] L. M. Roth, B. Lax, and S. Zwerdling, Phys. Rev. \textbf{114}, 90 (1959).\newline
[S5] O. B. O. Ly, Electron spin resonance in topological insulators: Theoretical study, Master's thesis, Université de Strasbourg (2014).\newline
[S6] H. Köhler and E. Wöchner, physica status solidi (b) \textbf{67}, 665 (1975).\newline
[S7] B. Fauqué \textit{et al.}, Phys. Rev. B \textbf{87}, 035133 (2013).\newline
[S8] H. Köhler and J. Hartmann, phys. status solidi (b) \textbf{63}, 171 (1974).\newline
[S9] D. Greenaway and G. Harbeke, J. Phys. Chem. Solids \textbf{26}, 1585 (1965).\newline
[S10] K. W. Post \textit{et al.}, Phys. Rev. B \textbf{88}, 075121 (2013).\newline
[S11] Y. Xia \textit{et al.}, Nature Phys. \textbf{5}, 398 (2009).\newline
[S12] Z.-H. Zhu \textit{et al.}, Phys. Rev. Lett. \textbf{107}, 186405 (2011).\newline
[S13] I. A. Nechaev \textit{et al.}, Phys. Rev. B \textbf{87}, 121111 (2013).\newline
[S14] O. V. Yazyev \textit{et al.}, Phys. Rev. B \textbf{85}, 161101 (2012).\newline
[S15] I. Aguilera \textit{et al.}, Phys. Rev. B \textbf{88}, 045206 (2013).\newline
[S16] A. M. Witowski \textit{et al.}, Phys. Rev. B \textbf{82}, 165305 (2010).\newline

\end{widetext}

\end{document}